\documentclass[aps,prb,twocolumn,english,showpacs,superscriptaddress,amssymb,amsfonts,nofootinbib]{revtex4-2}

\usepackage{graphicx}
\usepackage{amsmath}
\usepackage{braket}
\usepackage{tikz}
\usepackage{float}
\usetikzlibrary{positioning}
\usepackage{soul,xcolor}
\usepackage[export]{adjustbox}
\usepackage{placeins}
\usepackage[scr=boondox]{mathalpha}
\usepackage{ulem}
\usepackage{ragged2e}
\usepackage{amsthm}

\usepackage{epsfig}
\usepackage{amsfonts}
\usepackage{mathrsfs}
\usepackage{txfonts}
\usepackage{subfig}
\usepackage{braket}

\usepackage{xr}
\usepackage[linktocpage]{hyperref}
\hypersetup{colorlinks=true,citecolor=blue,linkcolor=blue, urlcolor=blue, breaklinks=true}

\newcommand{\Rom}[1]{\uppercase\expandafter{\romannumeral #1\relax}}

\newtheorem*{assumption*}{\assumptionnumber}
\providecommand{\assumptionnumber}{}


\theoremstyle{definition}

\theoremstyle{theorem}

\theoremstyle{corollary}

\theoremstyle{lemma}

\theoremstyle{Proposition}

\theoremstyle{definition}

\makeatletter

\newcommand{\Rmnum}[1]{\expandafter\@slowromancap\romannumeral #1@}
\makeatother
\newcommand{\imth}{\hspace{1pt}\mathrm{i}\hspace{1pt}}

\newcommand{\bea}{\begin{eqnarray}}
\newcommand{\eea}{\end{eqnarray}}
\newcommand{\bct}{\begin{center}}
\newcommand{\ect}{\end{center}}
\newcommand{\bpm}{\begin{pmatrix}}
\newcommand{\epm}{\end{pmatrix}}
\newcommand{\bal}{\begin{aligned}}
\newcommand{\eal}{\end{aligned}}

\newcommand{\trace}{\text{Tr}}

\begin{document}
\title{Spacetime symmetry indicators for two-dimensional Floquet topological insulators}
 \author{Shuangyuan Lu}
 \author{Yuan-Ming Lu}
 \affiliation{Department of Physics, The Ohio State University, Columbus, OH 43210, USA}

\date{\today}

\begin{abstract}
Floquet Topological Insulators (FTIs) in two spatial dimensions can exhibit anomalous chiral edge modes despite a fully localized bulk, captured by a new topological invariant other than the Chern number. In this work, we focus on $(2+1)D$ FTIs in symmetry class A with a spacetime symmetry known as $n$-fold time screw symmetry. We show that the bulk invariant can be partially determined by symmetry indicators associated with $n=2,3,4,6$ time screw symmetry, which only depend on the time evolution in high symmetry momenta. We derive simple formula that relates symmetry indicators to the bulk invariant, and demonstrate its validity in examples of FTI models. Our results can simplify the engineering of FTIs in real materials, and especially shed light on systems driven by circularly polarized light. 
\end{abstract}

\maketitle

\emph{Introduction}
Topological insulators\cite{Hasan_2010, Hasan_2011,Qi_2011} are gapped quantum phases of electrons that exhibit novel physical phenomena such as gapless boundary excitations\cite{Chang_2013, Liu_2016, He_2018, Zhang_2013, Li_2016}.  Experimentally, solid state materials that realize topological states have been extensively studied for decades\cite{Chen_2009, Ando_2013, Ko_nig_2007}. Theoretically, these states are best understood in weakly interacting systems in the framework of electronic band theory. Since the nontrivial topology of the ground states are protected by certain symmetries, lots of works are devoted to the systematic classifications of topological insulators protected by internal\cite{Schnyder_2008, Kitaev_2009,Chiu2016} and crystalline symmetries\cite{Slager_2012,Ando2015,Kruthoff_2017}. 

Symmetry protection for the nontrivial topology is only one side of the story. On the other hand, symmetry quantum numbers can not only constrain but also indicate the topology of the ground state, known as symmetry indicators \cite{Fu2007,Po_2017,Bradlyn2017,Po_2020,Elcoro2021} for topology. Symmetry indicators can be used to extract partial information of the topological invariant, hence significantly simpliying the discovery of topological materials. In fact, they are widely used in large-scale prediction and discovery of topological materials\cite{Tang_2019a,Zhang2019,Vergniory2019,Karaki2023,Xu2024}.

Some materials without a topological ground state, can realize nontrivial topology in their dynamics under a periodic driving force, and they are called Floquet topological insulators\cite{Oka2009, Lindner_2011,Harper2020} (FITs).  As a direct generalization of the spatial periodicity in crystals to the time direction, Floquet systems can be used to engineer topological states defined similarly in electronic bands\cite{Harper2020,Oka_2019,Rudner_2020}. Moreover, Floquet time evolution can also bring in new physics and realize dynamical topological properties intrinsically different from static systems \cite{Kitagawa_2010, Rudner_2013, Ladovrechis_2019, Huang_2020}. In these dynamical FTIs, topological edge states across the quasi-energy domain can appear in spite of trivial static topology in all Floquet bands.

The concept of symmetry indicator for topology also applies to Floquet systems\cite{Yu_2021}. Interestingly, symmetries are not restricted to global symmetries and space groups, but can be generalized to spacetime symmetries\cite{Xu_2018}. New physics can arise from the interplay of spacetime symmetry and topology in Floquet systems\cite{Peng_2019, Morimoto_2017, Chaudhary_2020}. 

In this paper, we study space-time symmetry indicators for FTIs of symmetry class A\cite{Roy_2017} in $(2+1)D$, also known as anomalous Floquet-Anderson insulators\cite{Rudner_2013,Titum2016}. We focus on  a special type of space-time symmetry known as the $n$-fold time screw symmetry\cite{Xu_2018}. We derive simple symmetry indicator formula for the topological invariant of these FTIs with $n=2,3,4,6$, which only depends on the Floquet evolution at certain high symmetry momenta. We further demonstrate the validity and power of the symmetry indicator formula in models of the FTIs. 


\emph{Floquet loop evolutions with time screw symmetry}
We first set up the problem of a Floquet time evolution with the time screw symmetry. We focus on $(2+1)D$ FTIs of non-interacting fermions in symmetry class A, and study the topology of their Floquet time evlution. The time-dependent Hamiltonian $\hat{H}(t)$ is defined on a two-dimensional (2D) lattice and it is periodic in both time and space directions:
\bea
&\hat{H}(t + T) = \hat{H}(t) \\
&\hat{T}_i \hat{H} \hat{T}_i^{-1} = \hat{H}
\eea
Here, $T$ is the time period and $\hat{T}_i$ are the discrete translation operators in two directions $i = x,y$ of the 2D system. Because of its spatial periodicity, the Hamiltonian can be written in momentum space as
\bea
\hat{H}(t) = \sum_{\mathbf{k},\alpha, \beta} \hat{c}^{\dagger \alpha}(\mathbf{k})H_{\alpha, \beta}(\mathbf{k}, t) \hat{c}^\beta(\mathbf{k}) 
\eea
where $\mathbf{k}$ labels quasi-momentum in the 1st Brillouin zone and $\alpha, \beta$ labels sublattice sites. Hereafter we omit index $\alpha, \beta$ and use $H(\mathbf{k}, t)$ to represent the Hamiltonian matrix. The unitary evolution operator is defined as 
\bea
 U(\mathbf{k},t)=\hat{\mathcal{T}} ~ \mathbf{exp}(-{\imth\int_0^tH(\mathbf{k},\tau)\mathbf{d\tau}})
\eea
where $\hat{\mathcal{T}}$ denotes the time ordered exponential. In this work we focus on Floquet loop evolution\cite{Roy_2017,Harper2020}
\bea
U(\mathbf{k},T)= U(\mathbf{k},0)=\mathbb{I}
\eea
which is sufficient to capture the FTI phases. Our results also apply to a generic Floquet evolution with gaps in its quasi-energy spectrum, as discussed in Appendix \ref{appendix.gap}.

The unitary loop evolution $U(\mathbf{k},t)$ can realize intrinsically dynamic FTIs with no static counterparts in 2D\cite{Rudner_2013}, also known as Anomalous Floquet-Anderson Insulators\cite{Titum2016}, characterized by the following winding number
\bea
\notag
&\nu[U] =\int \frac{dk^2dt}{24\pi^2}\ w[U]\\
&=\int \frac{dk^2dt}{24\pi^2} \trace(U^{-1}\partial_\mu UU^{-1}\partial_\nu UU^{-1}\partial_\rho U) \epsilon^{\mu \nu \rho}
\label{eq.winding}
\eea
where $\mu, \nu, \rho=x, y, t$ and $\epsilon^{\mu \nu \rho}$ is the anti-symmetric tensor. 

The above winding number is an integral function of the full spacetime unitary evolution $U(\mathbf{k},t)$. Our goal is to show that in the presence of proper symmetries, it can be indicated by only the unitary evolution at high symmetry momenta. Specifically we consider a spacetime symmetry known as the $n$-fold time screw symmetry\cite{Xu_2018}, which is a combination of $n$-fold spatial rotation $\hat{C}_n$ and a translation in the time direction:
\bea
\hat{H}(\hat{C}_n\mathbf{k},t) = \hat{C}_n \hat{H}(\mathbf{k},t-\frac{T}{n}) \hat{C}_n^{-1}
\eea
where $n=2, 3, 4, 6$ in $2D$ crystals. Consequently, the unitary evolution operator transforms under time screw symmetry as
\bea \label{eq.U_Cn}
\hat{U}(\hat{C}_n\mathbf{k},t)=\hat{C}_n \hat{U}(\mathbf{k},t-\frac{T}{n}) \hat{C}_n^{-1} \hat{U}(\hat{C}_n\mathbf{k},\frac{T}{n})
\eea
While a general momentum is not preserved by rotation $\hat{C}_n$, certain high-symmetry momenta ${p}$ in the 1st Brillouin Zone can be invariant under $\hat{C}_p\equiv(\hat{C}_n)^{n/n_p}$, up to reciprocal lattice vectors, where $n_p$ is a factor of $n$. 
\begin{figure}[h]
    \centering
    \includegraphics[width=\columnwidth]{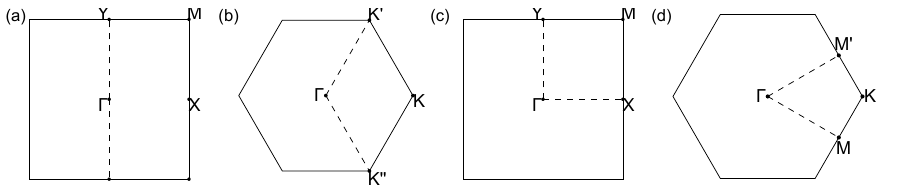}
    \caption{Brillouin Zone of lattices with $C_2,C_3,C_4,C_6$ symmetry with high-symmetry momentum points labeled. Dashed lines are used enclose regions of  symmetry independent momentum.}
    \label{fig.BZ}
\end{figure}
We label these high-symmetry momenta for $n=2,3,4,6$ lattices in Fig. \ref{fig.BZ}. In the $n=2$ case, momenta $\Gamma, X, Y, M$ are invariant under $\hat{C}_2$, so $n_\Gamma = n_M = n_X = n_Y = 2$. In the $n=3$ case, $\Gamma, K, K'$ are invariant under $\hat{C}_3$, so $n_\Gamma = n_K = n_{K^\prime} = 3$. In the $n=4$ case, $\Gamma, M$ are invariant under $\hat{C}_4$ and $X, Y$ are invariant under $\hat{C}_4^2$, so $n_\Gamma = n_M = 4, n_X = n_Y = 2$. In the $n=6$ case, $\Gamma$ is invariant under $\hat{C}_6$, $M$ is invariant under $\hat{C}_6^3$ and $K$ is invariant under $\hat{C}_6^2$, so $n_\Gamma = 6, n_M = 2, n_K = 3$. 

The general constraint of time screw symmetry on a high-symmetry momentum $p$ is
\bea \label{eq.U_Cp}
U(p,t)=C_p U(p,t-\frac{T}{n_p}) C_p^{-1} U(p, \frac{T}{n_p})
\eea
Here $C_p$ is a matrix representing the $p$-fold spatial rotation on momentum $p$ and it satisfies 
\bea \label{eq.C_p}
C_p^{n_p} = \mathbb{I}
\eea

\emph{Symmetry indicator formula} In the presence of $n$-fold time screw symmetry, can we obtain the winding number $\nu$ in Eq. (\ref{eq.winding}) using only the Floquet evolution at all high-symmetry momenta? We show a positive answer below, by deriving simple formula that indicates the bulk invariant $\nu$ based only on data at high-symmetry momenta. 

We first define the ``symmetry indicators'' from the loop evolution at high-symmetry momenta. Since ${C}_p$ is a $n_p$ fold symmetry satisfying Eq.\ref{eq.C_p}, its eigenvalues must be $n_p$-th roots of unity. Hence we can define a matrix ${S}_p$ as
\bea
\label{eq.S_p}
C_p \equiv\exp{(-2 \pi \imth{S}_p / n_p)}
\eea
whose eigenvalues are all integers. Note this equation does not define $S_p$ uniquely, and we will address this issue later. 

Next, this unitary loop evolution must satisfy
\bea
\notag
&U(p, T)=\prod_{i = 0}^{n_p - 1} C_p^i U(p, \frac{T}{n_P}) C_p^{-i} \\
&= \left( C_p^{-1} U(p, \frac{T}{n_p}) \right) ^{n_p} = \mathbb{I}
\eea
In the above product, terms with larger $i$ are multiplied on the l.h.s.. Therefore $C_p^{-1} U(p, \frac{T}{n_p})$ is another matrix whose eigenvalues are $n_p$-th roots of unity. Similarly we can define another matrix $\widetilde{H}_p$
\bea \label{eq.H_p}
\widetilde{U}_p = C_p^{-1} U(p, \frac{T}{n_p})\equiv \exp{(-2 \pi\imth\widetilde{H}_p / n_p)} 
\eea
whose eigenvalues are all integers. 

Armed with the two symmetry indicator matrices $S_p$ and $\tilde H_p$, we are ready to write down the symmetry indicator formula for bulk invariant $\nu$
\begin{align}
\label{eq.SI_2}
\nu_2[U]&=\frac12 \trace \Big ( ( \widetilde{H}_\Gamma^2+\widetilde{H}_M^2- \widetilde{H}_X^2-\widetilde{H}_Y^2) \\
\notag
&-(S_\Gamma^2+S_M^2-S_X^2-S_Y^2) \Big ) \quad\mod \ 2 \\ 
\label{eq.SI_3}
\nu_3[U]&=\frac12 \trace  \Big( (\widetilde{H}_\Gamma^2+\widetilde{H}_K^2-2 \widetilde{H}_{K'}^2)\\
\notag
&- (S_\Gamma^2+S_K^2-2 S_{K'}^2) \Big) \quad\mod \ 3\\
\label{eq.SI_4}
\nu_4[U]&=\frac12 \trace  \Big ( (\widetilde{H}_\Gamma^2+\widetilde{H}_M^2-2 \widetilde{H}_X^2) \\
\notag
&- (S_\Gamma^2+S_M^2-2 S_X^2) \Big) \quad\mod \ 4\\
\label{eq.SI_6}
\nu_6[U]&=\frac12 \trace  \Big( (\widetilde{H}_\Gamma^2+2 \widetilde{H}_K^2-3 \widetilde{H}_M^2) \\ 
\notag
&-  (S_\Gamma^2+2 S_K^2-3 S_M^2) \Big) \quad\mod \ 6
\end{align}
for $n=2,3,4,6$-fold time screw symmetry respectively. As shown above, the bulk invariant $\nu$ can be determined by the integer eigenvalues of symmetry indicator matrices $S_p$ and $\tilde H_p$ at high symmetry momenta $p$. 

There is an important subtlety though. The exponential functions in Eq. (\ref{eq.S_p}) and (\ref{eq.H_p}) do not uniquely define $S_p$ and $H(p)$, but only fix their integer eigenvalues modulo $n_p$. Fortunately, in $n_p=2,4,6$ cases, the l.h.s. of Eq. (\ref{eq.SI_2}), (\ref{eq.SI_4}), (\ref{eq.SI_6}) modulo $n$ remains invariant if one eigenvalue of $S_p$ (or $\tilde H_p$) changes by multiples of $n_p$. However, this is not the case for $n_p=3$ momenta, i.e. $\Gamma, K, K^\prime$ in Eq. (\ref{eq.SI_3}) and $K$ in (\ref{eq.SI_6}). In these cases, a change of eigenvalues by $n_p=3$ will change the l.h.s. by $\frac{3}{2}$ in the $n=3$ case and by $3$ in the $n=6$ case.


In these ambiguous cases, we need to use the rigorous definition of the symmetry indicator matrices $\tilde H_p$ beyond Eq. (\ref{eq.H_p}). We first introduce two unitary evolutions as a function of time $t$:
\bea
&\widetilde{U}_p(t) = \exp{(-i 2 \pi \widetilde{H}_pt/T)} \\
&\widetilde{U}_0(p, t)=\left\{\begin{array}{ll} 
U(p, 2t)\quad & 0 \leqslant t \leqslant \frac{T}{2n_p} \\
\exp{(i 2\pi S_p \frac{2t-\frac{T}{n_p}}{T})} U(p, \frac{T}{n_p})  & \frac{T}{2n_p} < t \leqslant \frac{T}{n_p}
\end{array}\right.~~~\label{eq.U_0}
\eea
Note that at time $t=T/n_p$, $\widetilde{U}_p(t=T/n_p)$ recovers $\widetilde{U}_p$ on the r.h.s. of Eq. (\ref{eq.H_p}), while $\widetilde{U}_0(p, t=T/n_p)$ recovers the l.h.s. of Eq. (\ref{eq.H_p}). Our requirement for $\tilde H_p$ is that the evolution $\widetilde{U}_0(p,t)$ can be continuously tuned into the evolution $\widetilde{U}_p(t)$ with the same fixed ending point at $t=T/n_p$, as guaranteed by Eq. (\ref{eq.H_p}). This means $\widetilde{U}_p(t)$ and $\widetilde{U}_0(p,t)$ must have the same $1D$ winding number:
\bea
\notag
\label{eq.1d_winding}
\frac{\imth}{2\pi} \int_0^{\frac{T}{n_p}} \mathbf{d}t \left( \widetilde{U}_0^{-1}(p, t) \partial_t \widetilde{U}_0(p, t) -  \widetilde{U}^{-1}_p(t)  \partial_t \widetilde{U}_p(t)\right) = 0\\
\eea
Since a shift of $\widetilde{H}_p$ eigenvalue by $2n_p = 6$ will not influence symmetry indicator formula (\ref{eq.SI_3}) and (\ref{eq.SI_6}), we can relax the above constraint by requiring Eq. (\ref{eq.1d_winding}) to be an even integer. Since the l.h.s. of Eq. (\ref{eq.1d_winding}) depends on both $S_p$ and $\tilde H_p$, we can choose an arbitrary $S_p$ satisfying Eq. (\ref{eq.S_p}), and any $\tilde H_p$ satisfying Eq. (\ref{eq.H_p}) and (\ref{eq.1d_winding}) can lead to a correct invariant $\nu\mod n$ in Eq. (\ref{eq.SI_3}) and (\ref{eq.SI_6}).

This concludes our main results of symmetry indicator formula. We refer readers to Appendix \ref{appendix.proof} for rigorous proof and relevant details. Note that similar symmetry indicator formula can be derived for $n$-fold spatial rotational symmetry, reducing the bulk invariant $\nu$ to $1D$ winding numbers at the high symmetry momenta, as discussed in Appendix \ref{appendix.rotation}.

\emph{Examples}
Next we use two examples to demonstrate the symmetry indicator formula for $(2+1)D$ FTI models in $n=4,6$ cases. 

First, We consider the checkerboard-lattice RLBL model introduced in Ref.\cite{Rudner_2013}. We choose parameters $\delta_{AB} = 0, J = 2\pi/T$ so that the model supports a FTI with $4$-fold time screw symmetry, given by time-dependent Hamiltonian:
\bea
H_n(\mathbf{k})= - J \left( e^{i\mathbf{b}_n \cdot \mathbf{k}_n} \sigma^+ + e^{-i\mathbf{b}_n \cdot \mathbf{k}_n} \sigma^-\right)
\eea
Here $n = 1, 2, 3, 4$ represents the $n$-th step of the driving protocol between time $\frac{(n-1)T}{4}$ to $\frac{nT}{4}$. Vector $\mathbf{b}_n$ represents the direction of hopping: $\mathbf{b}_1 = -\mathbf{b}_3 = (a, 0)$,  $\mathbf{b}_2 = -\mathbf{b}_4 = (0, a)$ where $a$ is the lattice constant. The time screw symmetry is represented by
\bea
&C_4 = 
\begin{pmatrix}
    1 & 0\\
    0 & 1
\end{pmatrix}\\
&H_{n+1}(k_x, k_y) = C_4 H_n(k_y, -k_x) C_4^{-1}
\eea
The relevant high-symmetry points are $\Gamma = (0, 0), X = (\frac{\pi}{2a}, \frac{\pi}{2a}), M = (0, \frac{\pi}{a})$. Under $C_4$ rotation, momentum $\Gamma$ is not changed, but $X, M$ will shifted by a reciprocal lattice vector $\mathbf{G}_1 = (\frac{\pi}{a}, \frac{\pi}{a})$ and we need to multiply $C_4$ by a unitary basis transformation $M$
\bea
G_1 = 
\begin{pmatrix}
    1 & 0 \\
    0 & e^{i \mathbf{b}_1 \cdot \mathbf{G}_1}
\end{pmatrix}
= 
\begin{pmatrix}
    1 & 0 \\
    0 & -1
\end{pmatrix}
\eea
Hence the $C_4$ symmetry at $\Gamma, M$ and $C_2$ symmetry at $X$ are implemented by
\bea
\notag
&C_\Gamma = C_4 = \mathbb{I} \\
&C_M = G_1^{-1} C_4 =  \sigma^z \\
\notag
&C_X = G_1^{-1} C_4 ^ 2 = \sigma ^z
\eea
From the Hamiltonian, we can obtain unitary evolution at these momenta
\bea
\notag
U(\Gamma, T/4) = -\imth\sigma^x \\
U(M, T/4) = -\imth\sigma^x \\
\notag
U(X, T/2) = -\imth\mathbb{I}
\eea

Matrices $S_p$ and $\widetilde{U}_p$ can be obtained using Eq. (\ref{eq.S_p}) and (\ref{eq.H_p}) and diagonalized, leading to the winding number based on Eq. (\ref{eq.SI_4}) as follows:
\bea
\notag
&\nu_4[U]=\frac{1}{2}\left[\left((-1)^2+1^2+0^2+2^2-2\times (1^2+0^2)\right) \right. \\
\notag
&-\left.\left(0^2+0^2+0^2+2^2-2\times(1^2+0^2)\right)\right] \\
&=1 \mod \ 4
\eea
This is consistence with actual winding number $\nu= 1$ obtained using Eq. (\ref{eq.winding}), manifested by the chiral edge mode in the open-boundary Floquet spectrum\cite{Rudner_2013}.

\begin{figure}[h]
    \centering
    \includegraphics[width=\columnwidth]{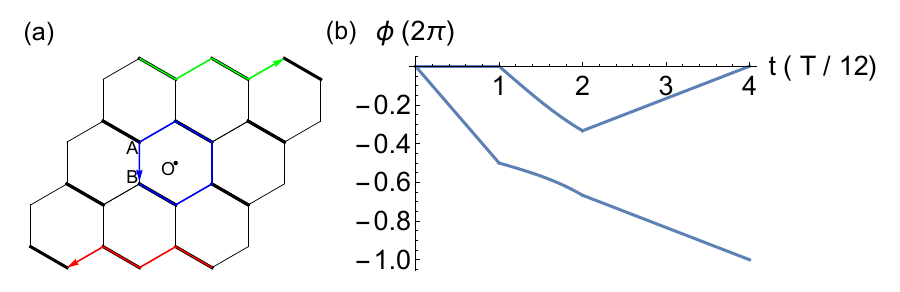}
    \caption{(a) Hexagonal lattice model with $6$-fold time screw symmetry. Distance between closest sublattice atoms $A$ and $B$ is a. Hopping is between the thick bonds in the first time period $T/6$ and rotate around point $O$ by $\pi / 3$ every time step in a counterclockwise manner. (b) Phase of eigenvalues of unitary evolution $\widetilde{U}_0(K)$ as defined in Eq. \ref{eq.U_0}. One eigenvalue goes from $1$ to $\exp{(-i 2\pi)}$.}
    \label{fig.eig}
\end{figure}

In the second example, we consider a honeycomb-lattice model with 6-fold time screw symmetry on hexagonal lattice, similar to Ref.\cite{Kitagawa_2010}. We divide the time period $T$ into $6$ steps. In the first
step ($0\leq t\leq T/6$), particles hop between two nearest atoms along the bold line in Fig. \ref{fig.eig} (a). The Hamiltonian is rotated around the hexagon center $O$ by $\pi/3$ after every step, manifesting the 6-fold time screw symmetry. The Hamiltonian in momentum space writes 
\bea
H_n(\mathbf{k}) =  J \left( e^{i\mathbf{b}_n \cdot \mathbf{k}_n} \sigma^+ + e^{-i\mathbf{b}_n \cdot \mathbf{k}_n} \sigma^-\right)
- J
\eea
where we used $J = \frac{3\pi}{T}$ and $\mathbf{b}_1 = \mathbf{b}_4 = (\sqrt{3}/2,-1/2)a$, $\mathbf{b}_2 = \mathbf{b}_5 = (-\sqrt{3}/2,-1/2)a$, $\mathbf{b}_3 = \mathbf{b}_6= (0, 1)a$, $a$ being the lattice constant. In this system, $C_6$ rotation is implemented by $C_6 = \sigma^x$. We can follow the previous example to obtain rotational symmetry matrix $C_p$, and evolution $U(p, T/n_p)$ of high symmetry momenta $\Gamma = (0, 0), M = (\sqrt{3} \pi /3a, -\pi / 3a), K = (4\sqrt{3} \pi /9a,0)$:
\bea
\notag
&C_\Gamma = \sigma^x \\ 
\notag
&C_M = G_1^{-1} C_6 ^ 3 = \begin{pmatrix}
    0 & \exp{(i 2\pi / 3)}  \\
     \exp{(-i 2\pi / 3)} & 0
\end{pmatrix}
\\
&C_K = G_1^{-1} C_6^2 = 
\begin{pmatrix}
    \exp{(i 2 \pi / 3)} & 0 \\
    0 & \exp{(-i 2 \pi / 3)}
\end{pmatrix}
\eea
\bea
\notag
& U(\Gamma, T/6) = \sigma^x \\
\notag
& U(M, T/2) = 
\begin{pmatrix}
    0 & -\exp{(-i \pi / 3)}  \\
    - \exp{(i \pi / 3)} & 0
\end{pmatrix}
\\
& U(K, T/3) = 
\begin{pmatrix}
    \exp{(i 2 \pi / 3)} & 0 \\
    0 & \exp{(-i 2 \pi / 3)}
\end{pmatrix}
\eea
Here, $G_1$ is the basis transformation between momenta differing by a reciprocal lattice vector $(\frac{2\sqrt{3} \pi}{3a}, -\frac{2\pi}{3a})$. 
\bea
G_1 = 
\begin{pmatrix}
    \exp{(-i 2 \pi / 3)} & 0 \\
    0 & \exp{(i 2 \pi / 3)}
\end{pmatrix}
\eea

With 6-fold time screw and $n_K=3$, we need to address the subtlety of defining $\widetilde{U}_K$. While Eq. (\ref{eq.H_p}) indicates that $e^{-2\pi\imth\tilde H_K/3}=\mathbb{I}$, we have to study $\widetilde{U}_0(K, t)$ defined in Eq. (\ref{eq.U_0}), and then use Eq. (\ref{eq.1d_winding}) to fix the ambiguity. By choosing $S_K = - \sigma^z$ from Eq. (\ref{eq.S_p}), we plot in Fig. \ref{fig.eig} the phases of eigenvalues of $\widetilde{U}_0(K, t)$ for $0\leq t\leq T/3$. 
It is clear that the phase of one eigenvalue winds from $0$ to $-2\pi$. According to Eq. (\ref{eq.1d_winding}) we can therefore fix $\widetilde{H}_K$ as
\bea
\widetilde{H}_K = 
\begin{pmatrix}
    0& 0 \\
    0& -3
\end{pmatrix}
\eea
The eigenvalue $-3$ enables unitary evolution $\widetilde{U}_0(p,t)$ to be continuously connected to $\widetilde{U}_p(t)$. Then we can obtain the bulk invariant (\ref{eq.SI_6}) as
\bea
\notag
&\nu_6[U]=\frac{1}{2}((0^2+0^2+2\times (0^2+(-3)^2)-3\times (0^2+0^2))\\
\notag
&-(3^2+0^2+2\times (1^2+(-1)^2)-3\times (1^2+0^2))) \\
&=-2 \mod \ 6
\eea
This is again consistent with bulk invariant $\nu=-2$ directly obtained from Eq. (\ref{eq.winding}), manifested in the two branches of chiral edge states on an open boundary.

One can also think of these two models as with $2$-fold and $3$-fold time screw symmetry respectively, by combining two consecutive steps of evolution into one. By using symmetry indicators formula (\ref{eq.SI_2}) and (\ref{eq.SI_3}), we can obtain $\nu_2=1\mod2$ and $\nu_3=-2\mod3$ respectively, again consistent with direct calculations. 

\emph{Summary and Outlook} 
In this paper, we considered $(2+1)D$ FTIs in symmetry class A in the presence of $n$-fold time screw symmetry. We derived symmetry indicator formula, Eq. (\ref{eq.SI_2})-(\ref{eq.SI_6}), which determines the bulk topological invariant modulo $n$ based only on Floquet evolutions at certain high-symmetry momenta in the 1st Brillouin zone. Remarkably, these symmetry indicators does not require full information of Floquet time evolutions at these momenta. The simplicity of the symmetry indicator formula allows a quick identification of nontrivial driven topology in Floquet systems, compared to a direct calculation of bulk topological invariant. 


Time screw symmetry can be naturally realized in rotational invariant systems driven by circularly polarized light, whose propagation direction is parallel to the rotation axis. AC Josephson junction is also a potential realization because the phase difference of order parameters is linear function of time. If the Hamiltonian is designed so that the space rotational symmetry shifts the phase of order parameter by $2\pi / n$, like in a $d$-wave superconductor, the Hamiltonian will be invariant under $n$-fold time screw symmetry. Our symmetry indicator formula can greatly reduce the amount of calculations in order to engineer FTIs in these systems, hence shedding light on future discovery of FTIs in driven solid state materials. 

\begin{acknowledgments}
This work is supported by Center for Emergent Materials at The Ohio State University, a National Science Foundation (NSF) MRSEC through NSF Award No. DMR-2011876. YML acknowledges partial support by grant NSF PHY-1748958 to the Kavli Institute for Theoretical Physics (KITP), and NSF PHY-2210452 to the Aspen Center for Physics.
\end{acknowledgments}

\bibliography{Symmetry_Indicator}

\clearpage
\onecolumngrid
\appendix

\section{A proof for symmetry indicator equations}
\label{appendix.proof}
In this section, we provide rigorous proofs of the main results presented in Eq. \ref{eq.SI_2}, \ref{eq.SI_3}, \ref{eq.SI_4}, \ref{eq.SI_6}. 

\subsection{Setting the stage: preliminary steps for a comprehensive proof}
In this subsection, we establish the basic setup, standardize our conventions and definitions, and derive several useful formulas.

We examine free Fermion models on a periodic $2D$ lattice, where sublattice orbits are denoted by the label $a$. The Hamiltonian exhibits a rotational symmetry within the plane. Assuming the center of rotation as the origin, we employ the following basis in the momentum space 
\begin{equation}\label{eq.basis}
    \left |\Psi_{a, \vec{k}} \right \rangle=\sum_{i} \mathbf{e}^{i\vec{k}\cdot \vec{x}_{a,i}}\left |\Psi_{a,i} \right \rangle
\end{equation}
where $i$ indexes each lattice unit cell, and $a$ labels different orbitals within a unit cell. The variable $\vec{x}_{ai}$ represents the displacement relative to the center of rotation. When the rotational symmetry is applied to this basis, it transforms orbital $a$ to orbital $b$ and momentum $\vec{k}$ to $\hat{C}_n(\vec{k})$
\bea
\hat{C}_n \left |\Psi_{a,\vec{k}} \right \rangle = \sum_{i} \mathbf{e}^{i\vec{k}\cdot (\hat{C}_n^{-1} \vec{x}_{b,i}) }\left |\Psi_{b,i} \right \rangle = \left |\Psi_{b, \hat{C}_n (\vec{k})} \right \rangle = \left( C_n \right) _{a, b}  \left|\Psi_{a, \hat{C}_n (\vec{k})} \right \rangle
\eea
where $C_n$ is a permutation matrix representing the action of $\hat{C}_n$. The primary advantage of this basis is that $C_n$ is independent of $\vec{k}$. However, a notable disadvantage arises from the basis discontinuity when the Brillouin zone is visualized as a torus. At the boundary of the Brillouin zone, two momentum points that correspond to the same quasi-momentum possess different bases. Specifically, for two momentum points $\vec{k}_1,\vec{k}_2$, they are related by $\vec{k}_1=\vec{k}_2+\vec{P}_j$, where $\vec{P}_j$ for $j=1,2$ are the unit vectors of the reciprocal lattice. 
The relationship between the two bases is expressed through the transformation defined by
\bea \label{eq.G}
    G(\vec{P_j})& \coloneqq \mathbf{diag}(\mathbf{e}^{i\vec{P_j}\cdot \vec{x}_a})
\eea
Here, $\mathbf{diag}$ indicates that the matrix is diagonal, with each diagonal element representing a phase factor $\mathbf{e}^{i\vec{P_j}\cdot \vec{x}_a}$ associated with orbital $a$. 
The transformation of basis and Hamiltonian due to $G(\vec{P}_j)$ can be described as follows:
\bea
    &\left |\Psi_{\vec{k}_1} \right \rangle = G(\vec{P}_j) \left |\Psi_{\vec{k}_2} \right \rangle\\
    &H(\vec{k}_2)=G(\vec{P}_j)H(\vec{k}_1)G^{-1}(\vec{P}_j) 
\eea

Next, we elucidate the relationship between the Hamiltonian, unitary evolution and symmetry. We express the time dependent Hamiltonian under the influence of time screw symmetry Eq. \ref{eq.U_Cn} explicitly as follows:
\bea
H(\vec{k}, t) = C_n H(\hat{C}_n^{-1}\vec{k},t-\frac{T}{n}) C_n^{-1}
\eea
This equation represents the constraint imposed by the time screw symmetry. Further more, the corresponding constraint on the unitary evolution matrix is captured by
\bea\label{eq.U_Cn_matrix}
    U(\vec{k},t)=C_n U(\hat{C}_n^{-1}\vec{k},t-\frac{T}{n}) C_n^{-1} U(\vec{k},\frac{T}{n})
\eea

 Here, we define our conventions for the Brillouin zone and the special high-symmetry points. Systems exhibiting $C_n$ symmetry permit the division of the Brillouin zone into $n$ distinct regions. Within one region, the Hamiltonian behaves independently ---- no momentum points are symmetrically related to each other within the same region, and any momentum in the Brillouin zone can be associated to a momentum in this region via symmetry transformations. Fig.\ref{fig.BZ} illustrates these regions, delineated by dashed lines. We will express the winding number only by the evolution $U$ within this region. We denote this $\frac{1}{n}$ portion of the Brillouin zone as $\mathcal{L}$. 
 
 In cases of $3$-, $4$- and $6$-fold symmetry, this region forms a  quadrilateral and with its vertices labeled counterclockwise from $\vec{k}=0$ as $\Gamma,X,M,Y$, as shown in Fig. \ref{fig.BZ/n}. In the $3$- and $6$-fold symmetric Brillouin Zone, the vertices $X,M,Y$ correspond to, respectively,  $K'',K,K'$ and $M,K,M'$, as depicted in Fig. \ref{fig.BZ}. For the sake of clarity, we unify these labels so that our proof explicit addresses all three cases. The $2$-fold symmetric case is slightly different, and the discussion below may not apply directly. We will explain why our result still holds for this case in Appendix \ref{appendix.C2}.

 The point $\Gamma$ is invariant under the $n$-fold rotation, while the point $M$ is invariant under the $n^\prime$-fold rotation up to a reciprocal lattice vector. The value of $n$ and $n^\prime$ for each symmetry case are as follows,
 \bea
\notag
&n=2,3,4,6\\
&n^\prime=2,3,4,3
\eea
Thus we can apply Eq. \ref{eq.U_Cn_matrix} to the points $\Gamma$ and $M$,
\bea
\notag
&U(\Gamma,t)=C_\Gamma U(\Gamma,t-\frac{T}{n}) C_\Gamma^{-1} U(\Gamma, \frac{T}{n}) \\
&U(M,t)=C_M U(M,t-\frac{T}{n^\prime}) C_M^{-1} U(M, \frac{T}{n^\prime})
\eea
Which matches Eq. \ref{eq.U_Cp} in the main text. Here, explicitly,
\bea
&C_\Gamma = C_n \\
&C_M = G C_n^{n/n^\prime}
\eea
$G$ represents the transformation of the basis that differs by a reciprocal lattice vector.
The point $X$ is also invariant under a rotation that combines the two previous rotations. Explicitly,
\bea
&U(x,t)=C(X) U(X,t-\frac{T}{n}-\frac{T}{n^\prime}) C(X)^{-1} U(X, \frac{T}{n} + \frac{T}{n^\prime})\\
&C(X )= C_M C_\Gamma =G C_n^{n/n^\prime + 1}
\eea
Here, $C(X)$ represents a rotation by the angle $\alpha_X 2\pi \equiv (\frac{1}{n}+\frac{1}{n^\prime})2 \pi$. ($C(X)$ is not defined this way for the $2$-fold case). We use $C(X)$ because it differs slightly difference from $C_X$ defined in the main text. In the $3$-fold case, $C(X)$ corresponds to a rotation by $4\pi / 3$, while $C_X$ represents the minimal invariant rotation of $2\pi / 3$.
For the four cases of $2$-, $3$-, $4$-, $6$-fold symmetries, we define the parameter $\alpha_X$ such that $\alpha_X 2\pi$ is the rotation angle under which $X$ remains invariant.
\bea
\alpha_X=\frac{1}{2},\frac{2}{3},\frac{1}{2},\frac{1}{2}
\eea
We also aim to define these rotations as a function of angle parameter $\alpha$. The definitions of $C_p(\alpha)$ (which do not apply to $2$-fold case) are as follows:
\begin{equation}\label{eq.C_X_alpha}
\begin{split}
\alpha_{\Gamma}&=\frac{1}{n} \quad  C_{\Gamma}(\alpha)= \exp{(-i 2\pi\alpha S_\Gamma )} \quad 0 \leqslant \alpha \leqslant \alpha_{\Gamma} \\
\alpha_{m}&=\frac{1}{n^{\prime}} \quad  C_{M}(\alpha)= \exp{(-i 2\pi \alpha S_M)} \quad 0 \leqslant \alpha \leqslant \alpha_{M}\\
\alpha_{X}&=\alpha_{\Gamma}+\alpha_{M} \quad C_{X}(\alpha)=\left\{\begin{array}{ll}
C_{\Gamma}(\alpha) & 0 \leqslant \alpha \leqslant \alpha_{\Gamma} \\
C_{M}(\alpha - \alpha_\Gamma) C_{\Gamma} & \alpha_{\Gamma}<\alpha \leqslant \alpha_{\Gamma}+\alpha_{M}
\end{array}\right.
\end{split}
\end{equation}
This definition is consistent with the earlier formulation:
\bea
C_p(0) = \mathbb{I}, \quad C_p(\frac{1}{n_p}) = C_p, \quad p = \Gamma, M
\eea
where $n_p$ corresponds to the symmetry of the point $p$, as defined in the main text.

The four edges of $\mathcal{L}$ are labeled as $u_s,s=0,\cdots, 3$ with their respective directions shown in Fig.\ref{fig.BZ/n}. We parametrize four edges using a scalar parameter $k\in [0,\pi]$ (Note that this $k$ is a scalar and distinct from the momentum $\vec{k}$). The values of $k=0$ and $k=\pi$ correspond to the tail and head of the arrow. We demonstrate this explicitly here
\begin{equation}
    \begin{split}
       u_0: \quad  \vec{k}_0&=\frac{\pi-k}{\pi} \vec{k}(\Gamma) +\frac{k}{\pi} \vec{k} (X)\\
       u_1: \quad  \vec{k}_1&=\frac{\pi-k}{\pi} \vec{k}(M) +\frac{k}{\pi} \vec{k} (X)\\
       u_2: \quad  \vec{k}_2&=\frac{\pi-k}{\pi} \vec{k}(M) +\frac{k}{\pi} \vec{k} (Y)\\
       u_3: \quad  \vec{k}_3&=\frac{\pi-k}{\pi} \vec{k}(\Gamma) +\frac{k}{\pi} \vec{k} (Y)\\
    \end{split}
\end{equation}
 
We will calculate the winding number, which is obtained as the integral of the winding number density. For convenience, We reproduce Eq. \ref{eq.winding} from the main text here.  
\bea
\notag
&\nu[U] =\int \frac{dk^2dt}{24\pi^2}\ w[U]\\
&=\int \frac{dk^2dt}{24\pi^2} \trace(U^{-1}\partial_\mu UU^{-1}\partial_\nu UU^{-1}\partial_\rho U) \epsilon^{\mu \nu \rho}
\label{eq.winding_number}
\eea
In Appendix \ref{appendix.basis}, we will demonstrate that the expression for the winding number given in Eq. \ref{eq.winding_number} is still valid even with our discontinuous basis as detailed in Eq. \ref{eq.basis}. As a preliminary step, we list some useful equations pertaining to the winding number density. Consider $U_1(\vec{k},t)$ and $U_2(\vec{k},t)$, which are two unitary matrices. Here, we will omit the arguments $(\vec{k},t)$ for simplicity when referring to $U_1$ and $U_2$. Then, we have the following expressions:
\begin{equation}\label{eq.U_12}
w [U_1 U_2^{-1}] =w[U_1]-w[U_2]-3 \trace \partial_\nu (U_1^{-1}\partial_\mu U_1 U_2^{-1}\partial_\rho U_2)
\end{equation}
\begin{equation}\label{eq.U_212}
\begin{split}
w[U_2U_1 U_2^{-1}]&=w[U_1]+3 \trace \partial_\nu (U_1\partial_\mu U_1^{-1} U_2^{-1}\partial_\rho U_2)\\
&-3 \trace \partial_\nu (U_1^{-1}\partial_\mu U_1 U_2^{-1}\partial_\rho U_2)\\
&-3 \trace \partial_\nu (U_1^{-1}U_2^{-1}\partial_\mu U_2 U_1 U_2^{-1}\partial_\rho U_2)
\end{split}
\end{equation}
These two equations will prove useful in our subsequent analysis.

\subsection{Rewriting the winding number integral using symmetry}
From this point, we begin to derive symmetry indicators from the expression of the winding number. Initially, we modify the form of the winding number by incorporating time screw symmetry.

\begin{figure}[h]
    \centering
    \begin{picture}(0,0)
        \put(-15, 165){\Large \textbf{(a)}}
    \end{picture}
    \includegraphics[width=0.35\columnwidth]{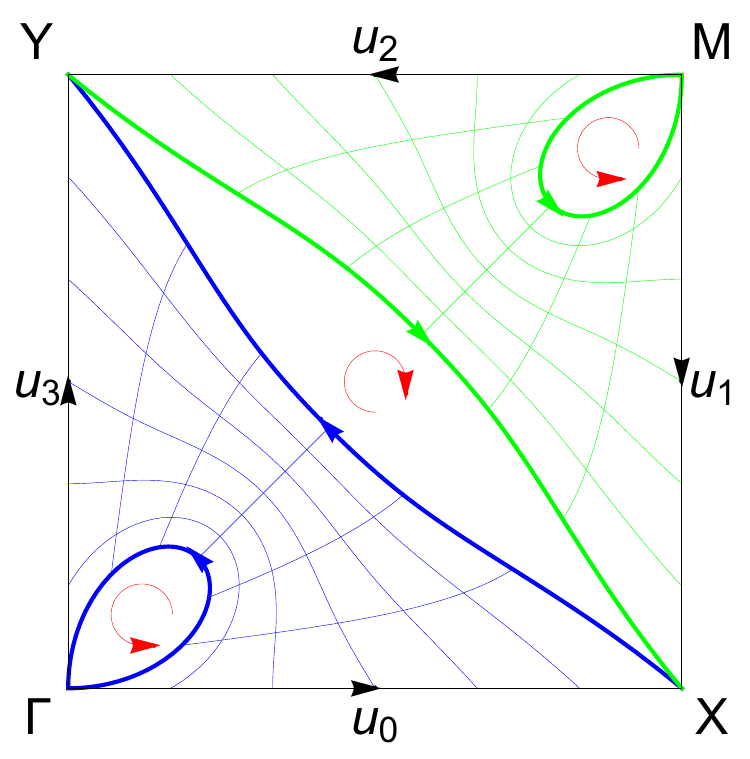}
    \hspace{0.3cm}
    \begin{picture}(0,0)
        \put(0, 150){\Large \textbf{(b)}}
    \end{picture}
    \includegraphics[width=0.25\columnwidth]{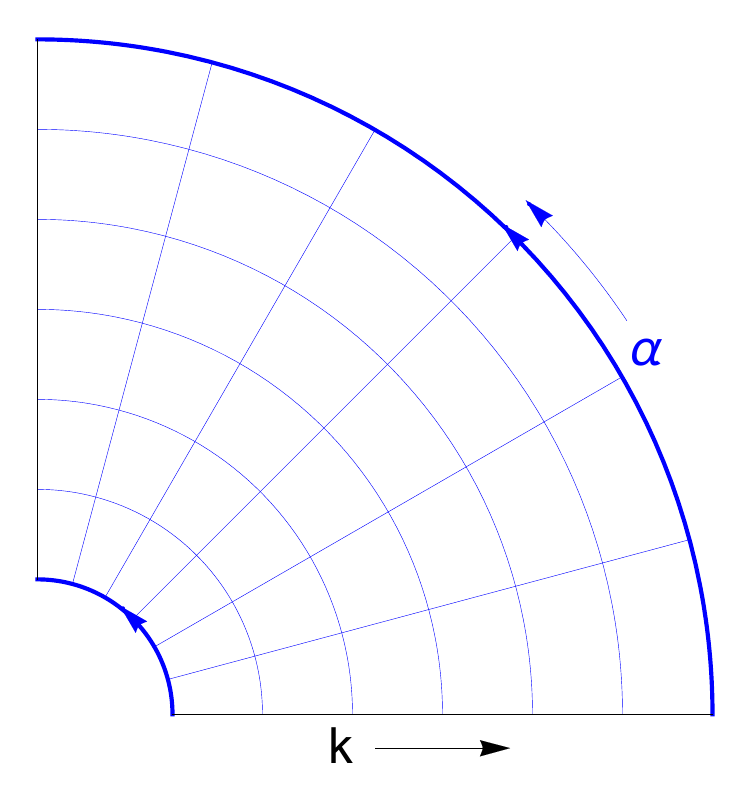}
    \begin{picture}(0,0)
        \put(0, 150){\Large \textbf{(c)}}
    \end{picture}
    \includegraphics[width=0.25\columnwidth]{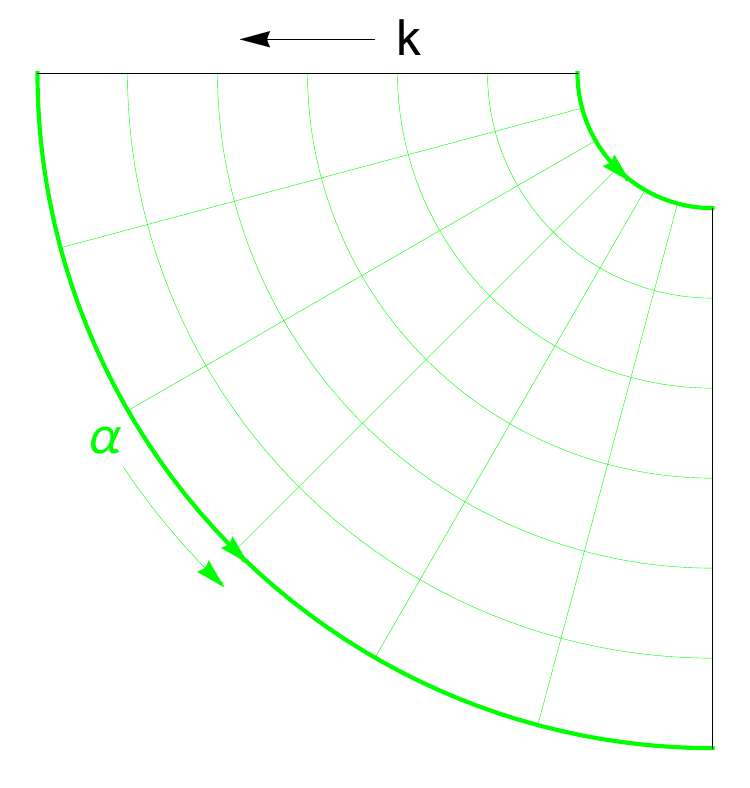}\\
    \begin{picture}(0,0)
        \put(-20, 90){\Large \textbf{(d)}}
    \end{picture}
    \includegraphics[width=0.2\columnwidth]{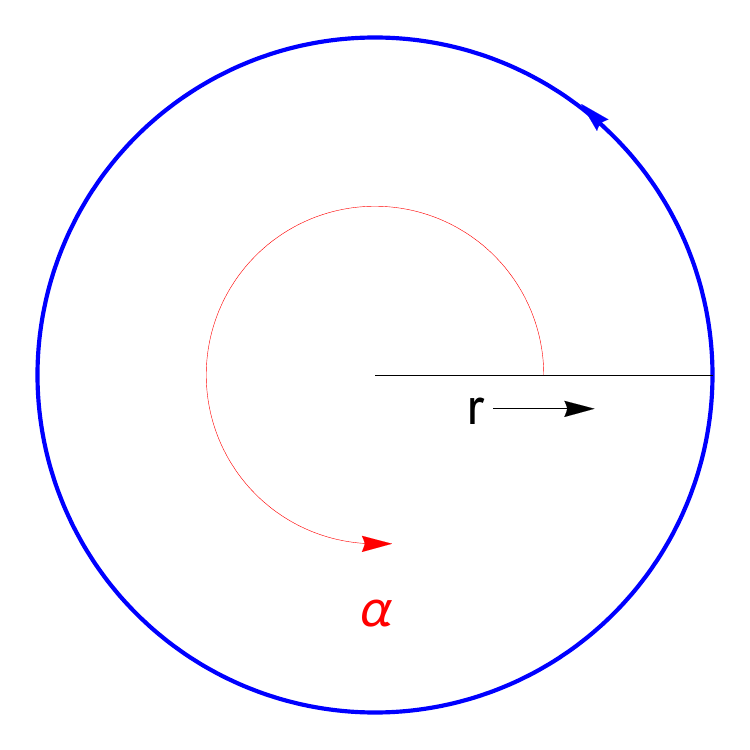}
    \hspace{0.5cm}
    \begin{picture}(0,0)
        \put(-20, 90){\Large \textbf{(e)}}
    \end{picture}
    \includegraphics[width=0.2\columnwidth]{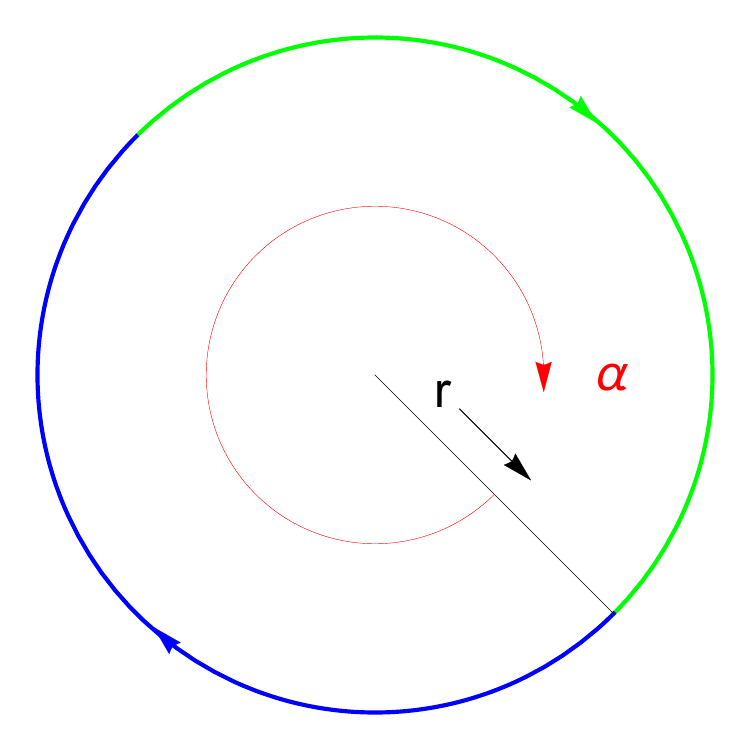}
    \hspace{0.5cm}
    \begin{picture}(0,0)
        \put(-20, 90){\Large \textbf{(f)}}
    \end{picture}
    \includegraphics[width=0.2\columnwidth]{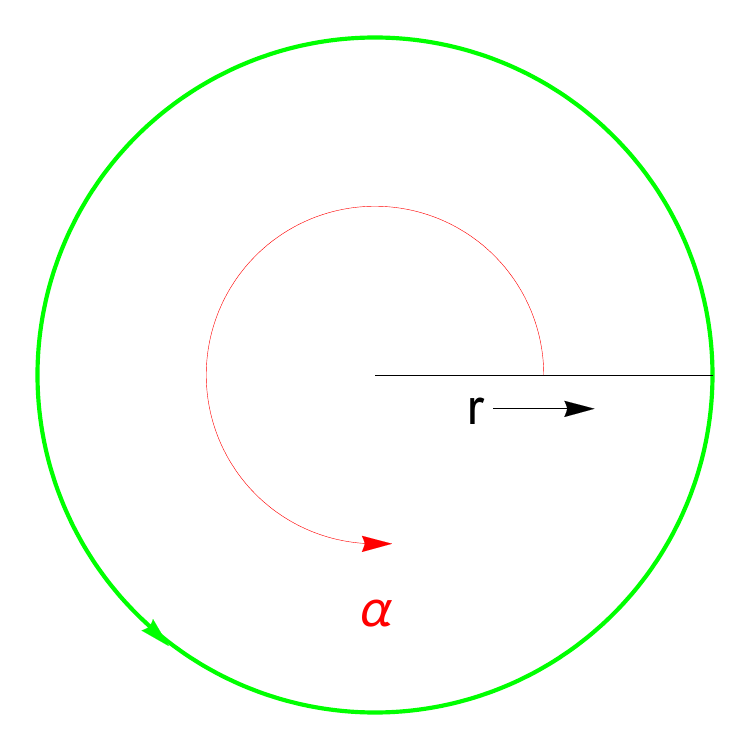}
    \caption{Construction of a new evolution $U^\prime$ in the region $\mathcal{L}$ using five segments. (a) The quadrilateral $\Gamma X M Y$ represents the region $\mathcal{L}$. (b) The first fan: The two radial edges are mapped to $\Gamma X$ and $\Gamma Y$. The inner arc is mapped to a loop at $\Gamma$. (c) The second fan: Two radial edges are mapped to $MY$ and $MX$. The inner arc is mapped to a loop at $M$. The two outer arcs of these two fans together form a loop at $XY$. (d) (e) (f) Three disks: Their circumferences are mapped to the three loops $\Gamma$, $XY$ and $M$ respectively. The evolution $U^\prime$ defined on the five segments can be equivalently interpreted as being defined within the region $\mathcal{L}$ by this mapping.}
    \label{fig.BZ/n}
\end{figure}
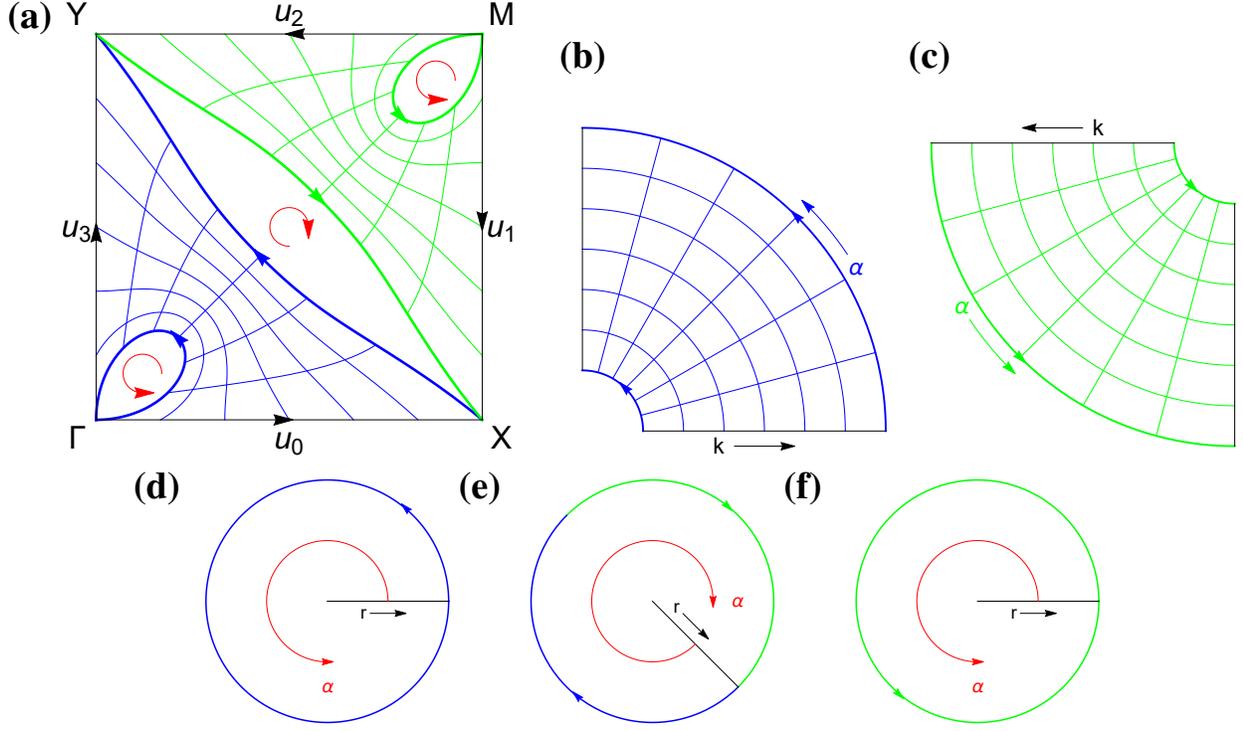

Starting from Eq. \ref{eq.winding_number}, we divide the integral into $n$ parts, each corresponding to a segment of the Brillouin zone associated with $\mathcal{L}$ through the rotation $\hat{C}_n^m$
\begin{equation}
\begin{split}
\nu[U]&=\sum_{m=0}^{n-1} \int_{\hat{C}_{n}^{m} \mathcal{L}} \frac{dk^2 dt}{24\pi^2}w[U(\vec{k}, t)]   \\
&=\sum_{m=0}^{n-1} \int_{ \mathcal{L}} \frac{dk^2 dt}{24\pi^2}w[U(\hat{C}_n^m\vec{k}, t+\frac{mT}{n})]   \\
\end{split}
\end{equation}
For each of these terms, by utilizing Eq. \ref{eq.U_Cn_matrix} and \ref{eq.U_12}, the unitary evolution can be expressed as the evolution within $\mathcal{L}$.
\begin{equation}
\begin{split}
  w[U(C_n^m\vec{k}, t+\frac{mT}{n})]
    &= w[C_n^m U(\vec{k},t) U^{-1}( \vec{k},-\frac{mT}{n})C_n^{-m}]\\
    &= w[U(\vec{k},t)] - \frac{3}{24\pi^2} \int \partial_\nu \trace (U(\vec{k},t)^{-1}\partial_\mu U(\vec{k},t) U^{-1}(\vec{k},-\frac{mT}{n})\partial _\rho U(\vec{k},-\frac{mT}{n}))
\end{split}
\end{equation}
where we have used the evolution for negative time as follows:
\bea
U^{-1}(\vec{k},-t)=\mathcal{T}e^{-i\int_{-t}^0H(\vec{k},\tau)\text{d}\tau},~~~t>0
\eea
Combining all $n$ terms, we obtain the following expression for the winding number $\nu[U]$
\begin{equation}
    \begin{split}
        \nu[U]=&n\int_{\mathcal{L}} w[U]\frac{dk^2 dt}{24\pi^2} \\
        &-3\sum_{m=0}^{n-1} \int_\mathcal{L}\frac{dk^2 dt}{24\pi^2}\ \partial_x \trace ( U^{-1}(\vec{k},t)\partial _t U(\vec{k},t)  U^{-1}(\vec{k},-\frac{mT}{n})\partial_y U(\vec{k},-\frac{mT}{n}))-(x\leftrightarrow y)\\
        =& n\int_{\mathcal{L}} w[U] \frac{dk^2 dt}{24\pi^2}\\
        &-\frac{3}{24\pi^2}\sum_{s=0}^3 (-)^s \sum_{m=0}^{n-1} \int_0^\pi dk\int dt \  \trace ( U^{-1}(\vec{k}_s ,t)\partial_t U(\vec{k}_s,t) U^{-1}(\vec{k}_s,-\frac{mT}{n})\partial_x U(\vec{k}_s,-\frac{mT}{n}))
    \end{split}
\end{equation}
Here, $u_s$ (for $s=0, 1, 2, 3$) represents four edges of the region $\mathcal{L}$ (as depicted in Fig. \ref{fig.BZ/n}). The sign associated with the edge $(-)^s$ depends on its orientation, with clockwise directions assigned a negative sign (-) and counterclockwise directions assigned a positive sign (+).
\begin{equation}
    \partial \mathcal{L}=\sum_{s=0}^3(-)^s u_s=u_0-u_1+u_2-u_3
\end{equation}

Consider the rotations that relate the edges $u_0$ and $u_3$, and $u_1$ and $u_2$ of the region $\mathcal{L}$. These relationship can be expressed mathematically as follows:
\begin{equation}
    \begin{split}
        \vec{k}_3&=\hat{C}_n \vec{k}_0\\
        \vec{k}_1&=\hat{C}_{n'} \vec{k}_2+\vec{P}
    \end{split}
\end{equation}
where $\vec P$ is a primitive vector of the reciprocal lattice. With these rotational relations established, We can now reformulate the expressions of the winding number as:
\begin{equation}\label{Eq.8}
    \begin{split}
        \nu[U]=&n\int_\mathcal{L} \frac{dk^2 dt}{24\pi^2}  w[U] \\
        &-\frac{3}{24\pi^2}\sum_{m=0}^{n-1} \int_0^\pi dk\int dt\  \trace \left( U^{-1}(\vec{k}_0,t)\partial _t U(\vec{k}_0,t) U^{-1}( \vec{k}_0,-\frac{mT}{n})\partial_x U( \vec{k}_0,-\frac{mT}{n}) \right)\\
        &+\frac{3}{24\pi^2}\sum_{m=0}^{n-1} \int_0^\pi dk\int dt\  \trace \left( C_n U( \vec{k}_0,-\frac{T}{n}) U(\vec{k}_0,t)^{-1}\partial _t U(\vec{k}_0,t) U^{-1}( \vec{k}_0,-\frac{T}{n}) C_n^{-1} \right. \\
        & \left. \quad\quad\quad\quad\quad\quad\quad\quad  U^{-1}(C_n \vec{k}_0,-\frac{mT}{n})\partial_k U(C_n \vec{k}_0,-\frac{mT}{n}) \right)\\ 
        &-\frac{3}{24\pi^2}\sum_{m=0}^{n-1} \int_0^\pi dk\int dt\  \trace \left( U^{-1}(\vec{k}_2,t)\partial _t U(\vec{k}_2,t)  U^{-1}(\vec{k}_2,-\frac{mT}{n})\partial_k U(\vec{k}_2,-\frac{mT}{n}) \right)\\
        &+\frac{3}{24\pi^2}\sum_{m=0}^{n-1} \int_0^\pi dk\int dt\  \trace \left( C_{n^\prime} U(\vec{k}_2,-\frac{T}{n}) U^{-1}(\vec{k}_2,t)\partial _t U(\vec{k}_2,t)  U^{-1}(\vec{k}_2,-\frac{T}{n})C_{n^\prime}^{-1} \right.\\
        &\left. \quad\quad\quad\quad\quad\quad\quad\quad  U^{-1}(\hat{C}_{n'} \vec{k}_2,-\frac{mT}{n})\partial_k U(\hat{C}_{n'}\vec{k}_2,-\frac{mT}{n}) \right)\\
    \end{split}
\end{equation}

In the last term, we use $\hat{C}_{n'} \vec{k}_2$ because the expressions for $\vec{k}$ and $\vec{k}+\vec{G}$ are equivalent except for a constant transformation of basis, leading to the same result. This identity is explicitly written as follows:
\begin{equation}\label{Eq.9}
    \begin{split}
        & \trace \left( U^{-1}(\vec{k}_1,t)\partial _t U(\vec{k}_1,t) U^{-1}( \vec{k}_1,-\frac{mT}{n})\partial_k U(\vec{k}_1,-\frac{mT}{n}) \right)\\
        =& \trace \left( U^{-1}(\hat{C}_{n'}\vec{k}_2 +\vec{P},t)\partial _t U(\hat{C}_{n'}\vec{k}_2 +\vec{P},t) U^{-1}( \hat{C}_{n'}\vec{k}_2 +\vec{P},-\frac{mT}{n})\partial_k U(\hat{C}_{n'}\vec{k}_2 +\vec{P},-\frac{mT}{n}) \right)\\
        =& \trace \left( U^{-1}(\hat{C}_{n'}\vec{k}_2,t)\partial _t U(\hat{C}_{n'}\vec{k}_2,t) U^{-1}( \hat{C}_{n'}\vec{k}_2,-\frac{mT}{n})\partial_k U(\hat{C}_{n'}\vec{k}_2,-\frac{mT}{n})\right)\\
    \end{split}
\end{equation}

Using the following two equations
\begin{equation}
    \begin{split}
U\left(\vec{k}_{0},-\frac{(m+1) T}{n}\right)&=C_{n}^{-1} U\left(C_{n} \vec{k},-\frac{m T}{n}\right) C_{n} U\left(\vec{k},-\frac{T}{n}\right) \\
U\left(\vec{k}_{2},-\left(\frac{m}{n}+\frac{1}{n^{\prime}}\right) T\right)&=C_{n^\prime}^{-1} U\left(C_{n^\prime} \vec{k}_{2},-\frac{m T}{n}\right) C_{n^\prime} U\left(\vec{k}_{2},-\frac{T}{n^{\prime}}\right)
    \end{split}    
\end{equation}
we substitute these into Eq. \ref{Eq.8}. By simplifying and canceling some terms, we obtain the expression for the winding number $\nu[U]$:
\begin{equation}\label{Eq.7}
    \begin{split}
        \nu[U]=n\bigg(&\int_{\mathcal{L}} \frac{dk^2 dt}{24\pi^2} w[U]\\
        &-\frac{3}{24\pi^2}\int_0^\pi dk \int_0 ^{2\pi} dt
        \trace \left( U^{-1}(\vec{k}_0,t)\partial_t U(\vec{k}_0,t)
         U^{-1}( \vec{k}_0,-\frac{T}{n}) \partial_k U(\vec{k}_0,-\frac{T}{n}) \right)\\
        &-\frac{3}{24\pi^2}\int_0^\pi dk \int_0 ^{2\pi} dt \trace \left ( U^{-1}(\vec{k}_2,t)\partial_t U(\vec{k}_2,t)
         U^{-1}(\vec{k}_2,-\frac{T}{n'}) \partial_k U(\vec{k}_2,-\frac{T}{n'}) \right) \bigg)\\
    \end{split}
\end{equation}

\subsection{Constructing unitary evolution $U^\prime$ : boundary consistency and integral calculation} \label{appendix.U_prime}
The second part of the proof involves designing another evolution $U^\prime$ that coincides with $U$ along the boundary of $\mathcal{L}$. As illustrated in Fig \ref{fig.BZ/n}, we partition the quadrilateral $\mathcal{L}$ into five distinct regions. The first two regions, depicted as blue and green fans, are parameterized by $k$ and a angle $\alpha$. The remaining three regions, shaped as disks, are parameterized by a radius $r$ and an angle $\alpha$. Each region corresponds to a specific part of the quadrilateral, as clearly indicated in Fig. \ref{fig.BZ/n}. Our task now is to define a new evolution $U^\prime$, demonstrate its continuity across the common boundaries of these five pieces, and calculate the integral of the winding number density of $U^\prime$.

We first define $U^\prime$ for the two fans as follows:
\begin{equation}\label{Eq.10}
U_{s}^{\prime}(k, \alpha, t) \coloneqq C_s(\alpha) U(\vec{k}_{s}, t-\alpha T) U^{-1}(\vec{k}_{s},-\alpha T) C_s^{-1}(\alpha)
\end{equation}
where $s=0$ and $s=2$ correspond to the blue and green parts, respectively, in Fig. \ref{fig.BZ/n}. The range of the parameter $\alpha$ and the definition of $C(\alpha)$ for the two cases are as follows:
\begin{equation}\begin{array}{l}
\alpha \in\left[0, \alpha_{s}\right] \quad k \in[0, \pi] \\
s=0 \quad \alpha_{s}=\frac{1}{n} \equiv \alpha_{\Gamma} \quad C_s(\alpha)=C_\Gamma(\alpha) \\
s=2 \quad \alpha_{s}=\frac{1}{n^{\prime}} \equiv \alpha_{M} \quad C_s(\alpha) = C_M(\alpha)
\end{array}\end{equation}

It is straightforward to verify that the definition of $U^\prime$ is consistent with the original evolution $U$ along the boundary of the quadrilateral region $\Gamma XMY$. Explicitly, we have
\bea
\notag
&U^\prime_s(k, 0, t) = U(\vec{k}_s, t) \\
\notag
&U^\prime_0(k, \alpha_\Gamma, t) = C_\Gamma U(\vec{k}_0, t - \alpha_\Gamma T) U^{-1}(\vec{k}_0, \alpha_\Gamma T) C_\Gamma^{-1} = U(\vec{k}_1, t) \\
&U^\prime_2(k, \alpha_M, t) = C_M U(\vec{k}_2, t - \alpha_M T) U^{-1}(\vec{k}_2, \alpha_M T) C_M^{-1} = U(\vec{k}_3, t) 
\eea

To calculate the density of winding number for the new evolution, we utilize the equations specified in Eq. \ref{eq.U_12} and \ref{eq.U_212}. Given that the index $\nu$ cannot be time $t$ due to the absence of a temporal boundary, and that $C_s(\alpha)$ only depends on the rotational angle $\alpha$, we formulate the contributions to the winding number as follows:
\begin{equation}\label{Eq.11}
\begin{split}
&\frac{1}{24\pi^2} \int_0^{2\pi} dt \int_0^\pi dk \int_0^{\alpha_s} d\alpha ~  w\left[U_{s}^{\prime}(k, \alpha, t)\right]\\
&=3 \int \frac{dtdkd\alpha}{24\pi^2} \trace \partial_{k}((U(\vec{k}_s, t-\alpha T) U^{-1}(\vec{k}_{s},-\alpha T)) \partial_{t}(U(\vec{k}_{s},-\alpha T) U^{-1}(\vec{k}_{s}, t-\alpha T)) {C}_s^{-1}(\alpha) \partial_{\alpha} C_s(\alpha))\\
&-3 \int \frac{dtdkd\alpha}{24\pi^2} \trace \partial_{k}((U(\vec{k}_{s},-\alpha T) U^{-1}(\vec{k}_{s}, t-\alpha T)) \partial_{t}(U(\vec{k}_{s}, t-\alpha T)) U^{-1}(\vec{k}_{s},-\alpha T))  C_s^{-1}(\alpha) \partial_{\alpha} C_s(\alpha)) \\
&-3 \int \frac{dtdkd\alpha}{24\pi^2} \trace \partial_{k}(U^{-1}(\vec{k}_{s}, t-\alpha T) \partial_{t} U(\vec{k}_{s}, t-\alpha T) U^{-1}(\vec{k}_{s},-\alpha T) \partial_\alpha U(\vec{k}_{s},-\alpha T)) \\
&-3 \int \frac{dtdkd\alpha}{24\pi^2} \trace \partial_{\alpha}(U^{-1}(\vec{k}_{s}, t-\alpha T) \partial_{t} U(\vec{k}_{s}, t-\alpha T) U^{-1}(\vec{k}_{s},-\alpha T) \partial_{k} U(\vec{k}_{s},-\alpha T))
\end{split}
\end{equation}
The upper and lower bounds are only written explicitly in the first line for brevity. We only have boundary terms. The terms on the two radial boundaries will accurately cancel boundary terms in Eq. {\ref{Eq.7}}. The terms on the arcs depend solely on the original evolution $U(\vec{k}, t)$ at high-symmetry points. Thus, we define a function $A(p, C(\alpha))$ for points $p = \Gamma,X,M$ and $C(\alpha) = C_\Gamma(\alpha), C_M(\alpha), C_X(\alpha)$
\begin{equation}
\begin{split}
A(p, C(\alpha)) &\equiv \frac{3}{24\pi^2} \int_0^{\alpha_C} d \alpha d t  \trace\\
&\left(U(\vec{k}(p), t-\alpha T) U^{-1}(\vec{k}(p),-\alpha T)\right) \partial_{t}\left(U(\vec{k}(p),-\alpha T) U^{-1}(\vec{k}(p), t-\alpha T)\right) C^{-1}(\alpha) \partial_{\alpha} C(\alpha)\\
&-\left(U(\vec{k}(p)-\alpha T) U^{-1}(\vec{k}(p), t-\alpha T)\right) \partial_{t}\left(U(\vec{k}(p), t-\alpha T) U^{-1}(\vec{k}(p),-\alpha T)\right) C^{-1}(\alpha) \partial_{\alpha} C(\alpha) \\
&-U^{-1}(\bar{k}(p), t-\alpha T) \partial_{t} U(\vec{k}(p), t-\alpha T) U^{-1}(\vec{k}(p),-\alpha T) \partial_{\alpha} U(\vec{k}(p),-\alpha T).
\end{split}
\end{equation}
Here, the upper limit $\alpha_C$ indicates the upper bound of interval for function $C(\alpha)$. It is important to note that when this function is used, the two parameters $p$ and $C(\alpha)$ might not match, implying a scenario where $C(\alpha) = C_{p^\prime}(\alpha)$ while $p \neq p^\prime$.

With the help of the definition of $A(p, C(\alpha))$, we neatly express the integral as follows:
\begin{equation}
    \begin{split}
        &\sum_{s=0,2}\int \frac{dtdkd\alpha}{24\pi^2} w\left[U_{s}^{\prime}(\vec{k}_{s}, \alpha, t)\right]\\
         =& -A\left(\Gamma, C_\Gamma(\alpha)\right)-A\left(M, C_M(\alpha)\right)+A\left(X, C_\Gamma(\alpha) \right)+A\left(Y , C_M(\alpha) \right)\\
        &+\sum_{s=0,2} \frac{3}{24\pi^2}\int_0^\pi dk \int_0^{2\pi} dt \trace (U^{-1}(\vec{k}_{s}, t) \partial_{t} U(\vec{k}_{s}, t) U^{-1}(\vec{k}_{s},-\alpha_0 T) \partial_{k} U(\vec{k}_{s},-\alpha_0 T))\\
    \end{split}
\end{equation}

We observe that the last boundary term is identical to that in Eq. \ref{Eq.7}. The other terms can be further summarized as functions $A$ at three points. For points $\Gamma$ and $M$, the expressions are already in their simplest form. For $X$, two functions $A$ can be further combined into one:
\begin{equation}
\begin{split}
& A\left(X, C_{\Gamma}(\alpha)\right)+A\left(Y, C_{M}(\alpha)\right) \\
=& A\left(X , C_{\Gamma}(\alpha)\right)+
\frac{3}{24\pi^2} \int_0^{\alpha_{M}} d \alpha d t  \trace\\
&\left(U(\vec{k}(X), t-(\alpha+\alpha_\Gamma)T) U^{-1}(\vec{k}(X),-(\alpha+\alpha_\Gamma)T)\right) \partial_{t}\left(U(\vec{k}(X),-(\alpha+\alpha_\Gamma)T) U^{-1}(\vec{k}(X), t-(\alpha+\alpha_\Gamma)T)\right)\\
&\times C_n^{-1}({MC_{n^\prime}})^{-1}(\alpha) \partial_{\alpha} ({MC_{n^\prime}})(\alpha)C_n\\
-&\left(U(\vec{k}(X)-(\alpha+\alpha_\Gamma)T) U^{-1}(\vec{k}(X), t-(\alpha+\alpha_\Gamma)T)\right) \partial_{t}\left(U(\vec{k}(X), t-(\alpha+\alpha_\Gamma)T) U^{-1}(\vec{k}(X),-(\alpha+\alpha_\Gamma)T)\right)\\
&\times ({MC_{n^\prime}})(\alpha)C_n\\
-&U^{-1}(\vec{k}(X), t-(\alpha+\alpha_\Gamma)T) \partial_{t} U(\vec{k}(X), t-(\alpha+\alpha_\Gamma)T) U^{-1}(\vec{k}(X),-(\alpha+\alpha_\Gamma)T) \partial_{\alpha} U(\vec{k}(X),-(\alpha+\alpha_\Gamma)T).\\
=& A\left(X , C_{X}(\alpha)\right)
\end{split}
\end{equation}

We thus obtain the integral of the first two regions (the two fans) as follows:
\begin{equation}\label{Eq.B1}
    \begin{split}
       B_{fans}= &\sum_{s=0,2}\int \frac{dtdkd\alpha}{24\pi^2} w\left[U_{s}^{\prime}(\vec{k}_{s},  \alpha, t)\right]\\
         =& -A\left(\Gamma, C_\Gamma(\alpha)\right)-A\left(M, M C_M(\alpha) \right)+A\left(X, C_{X}(\alpha) \right)\\
        &+\sum_{s=0,2} \frac{3}{24\pi^2}\int_0^\pi dk \int dt (U^{-1}(\vec{k}_{s}, t) \partial_{t} U(\vec{k}_{s}, t) U^{-1}(\vec{k}_{s},-\alpha_0 T) \partial_{k} U(\vec{k}_{s},-\alpha_0 T))\\
    \end{split}
\end{equation}

In addition to the two fans, there are there disks contributing to the remainder of the integral. To calculate this, we need to define three separate evolution matrices $U^\prime _p(r, \alpha, t)$, for each disk respectively. First, we set $U^\prime _p(r, \alpha = 0, t)$ as a continuous function of $r$ and $t$. On the boundary $r=0$ or $r=1$, we need the boundary condition as follows:
\bea\label{eq.tuning}
\notag
&U^\prime_p(r=1,0, t)=U(\vec{k}(p),t)\\
\notag
&U^\prime_p(r=0,0, t)=\mathbf{exp}(-i 2\pi H^\prime (p) t/T)\\
&H^\prime(p)=C_p(\alpha_p) H^\prime(p) C_p^{-1}(\alpha_p)
\eea
At $r=1$, the evolution corresponds to the original evolution at that point, and at $r=0$, it becomes the evolution of a time-independent Hamiltonian $H^\prime(p)$. Furthermore, this Hamiltonian $H^\prime(p)$ commutes with the symmetry operator $C_p(\alpha)$. The existence of such an evolution matrix $U^\prime _p(r, \alpha = 0, t)$ will be discussed in Appendix \ref{appendix.tuning}. 

Additionally, $U^\prime _p(r, \alpha = 0, t)$ must adhere to the following boundary condition at $t = T$:
\begin{equation} \label{eq.constraint}
    U_p^\prime(r,0, T)=\mathbb{I}
\end{equation}

Next, we introduce a new matrix $C_p^\prime(r, \alpha)$ that continuously depends on $r$ for points $X, \Gamma$ and $M$.
Specifically, for point $X$, we define the matrix $C_X^\prime(r, \alpha)$ such that 
\begin{equation}\label{Eq.C_X_prime}
    \begin{split}
        C_X^\prime(r=1, \alpha)&=C_X(\alpha)\\
        C_X^\prime(r=0, \alpha)&=\mathbf{exp}(-i 2\pi S_X \alpha)
    \end{split}
\end{equation}
Where $S_X$ is defined in the main text.
One requirement when changing $r$ is maintaining a boundary condition: 
\bea\label{eq.C_X_prime_condition}
C_X^\prime (r, \alpha_X) = C_X(\alpha_X)
\eea
The existence of this definition will also be proved in Appendix \ref{subsection.3}.

For points $\Gamma$ and $M$, The matrix $C_p^\prime(r, \alpha)$ is defined to be independent of the parameter $r$. The definition is formally expressed as follows:
\bea
C_p^\prime(r, \alpha) \coloneqq C_p(\alpha) = \mathbf{exp}(-i 2\pi S_p \alpha), \quad p = \Gamma, M
\eea

Equipped with the prior definitions of $U^\prime _p(r, \alpha=0, t)$ and $C_p^\prime(r, \alpha)$, the general definition of the new evolution matrix $U^\prime(r, \alpha, t)$ at any given $\alpha$ is given by:
\begin{equation}
U^\prime_{p}(r, \alpha, t) \coloneqq C^\prime_{p}(r, \alpha) U_p^\prime(r, 0, t-\alpha) U_p^{\prime -1}(r, 0, -\alpha) C_{p}^{\prime -1}(r, \alpha)
\end{equation}

This newly defined $U_p^\prime (r, \alpha, t)$ must satisfy specific boundary conditions. When $r = 1$, it is straightforward to check that this matrix is the same as matrix $U_s^\prime(k, \alpha, t)$ with $k = 0$ or $k = \pi$. Specifically, the relationships are given by:
\bea \label{eq.boundary_condition}
\notag
& U_\Gamma^\prime (r = 1, \alpha, t) = U_0^\prime(k = 0, \alpha, t) \\
\notag
& U_M^\prime (r = 1, \alpha, t) = U_2^\prime(k = 0, \alpha, t)\\
\notag
& U_X^\prime(r=1, \alpha, t) = \left\{
\begin{array}{ll}
U_0^\prime(k=\pi, \alpha, t) & 0 \leqslant \alpha \leqslant \alpha_{\Gamma} \\
U_2^ \prime (k=\pi, \alpha + \alpha_\Gamma , t)  & \alpha_{\Gamma}<\alpha \leqslant \alpha_{\Gamma}+\alpha_{M}
\end{array}\right.
\eea
On the other side, since $H^\prime(p)$ commutes with $C_p$ and therefore with $S(p)$, the evolution $U^\prime(t,r=0,\alpha)$ is independent of $\alpha$ for a given time $t$, thus representing a single point at the center of the disk (satisfying the boundary condition for $r=0$). 

With the evolution $U^\prime_{p}(r, \alpha, t)$ well defined, we start to calculate the integral using Eq. \ref{eq.U_12}, \ref{eq.U_212}.
\begin{equation}
    \begin{split}
    & \frac{1}{24\pi^2} \int_0^{2\pi} dt\int_0^1 dr \int_0^{\alpha_p} d\alpha \quad w[U^\prime_p(r,\alpha, t)]\\
        =&\quad \int \frac{dtdrd\alpha}{24\pi^2}  3\trace \partial_{\nu} \left[ \left(U_p^{\prime}(r, 0,t-\alpha) U_p^{\prime-1}(r, 0, -\alpha)\right) \partial_{\mu}\left(U_p^{\prime}(r , 0, -\alpha) U_p^{\prime-1}(r, 0, t-\alpha)\right) C_{p}^{\prime-1}(r, \alpha) \partial_{\rho} C^\prime_{p}(r, \alpha) \right]\\
        &-\int \frac{dtdrd\alpha}{24\pi^2} 3\trace \partial_\nu \left [ \left(U_p^{\prime}(r, 0, -\alpha) U_p^{\prime-1}(r, 0, t-\alpha)\right) \partial_{\mu}\left(U_p^{\prime}(r, 0, t-\alpha) U_p^{\prime-1}(r, 0, -\alpha)\right) C_{p}^{\prime-1}(r, \alpha) \partial_{\rho} C^\prime_{p}(r, \alpha) \right]\\
        & -\int \frac{dtdrd\alpha}{24\pi^2} 3\trace \partial_\nu \left [ U_p^\prime(r, 0, -\alpha) U_p^{\prime -1}(r, 0, t-\alpha) C_{p}^{\prime-1}(r, \alpha) \partial_{\mu} C^\prime_{p}(r, \alpha) U_p^\prime(r, 0, t-\alpha) U_p^\prime(r, 0, -\alpha)^{-1} C_{p}^{\prime-1}(r, \alpha) \partial_{\rho} C^\prime_{p}(r, \alpha)  \right]\\        
        &-\int \frac{dtdrd\alpha}{24\pi^2} 3\trace \partial \nu \left[ U_p^{\prime-1}(r, 0, t-\alpha) \partial_{\mu} U_p^{\prime}(r, 0, t-\alpha) U_p^{\prime-1}(r , 0, -\alpha) \partial_{\rho} U_p^{\prime}(r, 0, -\alpha)\right]
\end{split}
\end{equation}
Among these boundary terms, the index $\nu$ can not be time $t$ due to the absence of a boundary in the time direction. Additionally, $C^\prime_p(r, \alpha)$ does not depend on $t$. Thus, the second-to-last line of the calculation is zero because none of the indices $\nu, \mu, \rho$ can be $t$. Given this understanding, the only non-zero terms are as follows:
\begin{equation}\label{Eq.6}
    \begin{split}    
        =& -\int \frac{3dtdr}{24\pi^2} \trace \left(U_p^{\prime}(r, 0, t-\alpha_p) U_p^{\prime-1}(r, 0, -\alpha_p)\right) \partial_{t}\left(U_p^{\prime}(r , 0, -\alpha_p) U_p^{\prime-1}(r, 0, t-\alpha_p)\right) C_{p}^{\prime-1}(r, \alpha_p) \partial_{r} C^\prime_{p}(r, \alpha_p)\\
        &+\int \frac{3dtdr}{24\pi^2} \trace \left(U_p^{\prime}(r, 0, -\alpha_p) U_p^{\prime-1}(r, 0, t-\alpha_p)\right) \partial_{t}\left(U_p^{\prime}(r, 0, t-\alpha_p) U_p^{\prime-1}(r, 0, -\alpha_p)\right) C_{p}^{\prime-1}(r, \alpha_p) \partial_{r} C^\prime_{p}(r, \alpha_p)\\
        &+\int \frac{3dtdr}{24\pi^2} \trace \left(U_p^{\prime-1}(r, 0, t-\alpha_p) \partial_{t} U_p^{\prime}(r, 0, t-\alpha_p) U_p^{\prime-1}(r , 0, -\alpha_p) \partial_{r} U_p^{\prime}(r, 0, -\alpha_p)\right)\\
        + &\quad \int \frac{3d\alpha dt}{24\pi^2} \trace \left(U_p^{\prime}(1, 0, t-\alpha) U_p^{\prime-1}(1, 0, -\alpha)\right) \partial_{t}\left(U_p^{\prime}(1 , 0, -\alpha) U_p^{\prime-1}(1, 0, t-\alpha)\right) C_{p}^{\prime-1}(1, \alpha) \partial_{\alpha} C^\prime_{p}(1, \alpha)\\
        &-\int \frac{3d\alpha dt}{24\pi^2} \trace \left(U_p^{\prime}(1, 0, -\alpha) U_p^{\prime-1}(1, 0, t-\alpha)\right) \partial_{t}\left(U_p^{\prime}(1, 0, t-\alpha) U_p^{\prime-1}(1, 0, -\alpha)\right) C_{p}^{\prime-1}(1, \alpha) \partial_{\alpha} C^\prime_{p}(1, \alpha)\\
        &-\int \frac{3d\alpha dt}{24\pi^2} \trace\left(U_p^{\prime-1}(1, 0, t-\alpha) \partial_{t} U_p^{\prime}(1, 0, t-\alpha) U_p^{\prime-1}(1, 0, -\alpha) \partial_{\alpha} U_p^{\prime}(1, 0, -\alpha)\right)\\
        - &\quad \int \frac{3d\alpha dt}{24\pi^2} \trace \left(U_p^{\prime}(0, 0, t-\alpha) U_p^{\prime-1}(0, 0, -\alpha)\right) \partial_{t}\left(U_p^{\prime}(1 , 0, -\alpha) U_p^{\prime-1}(0, 0, t-\alpha)\right) C_{p}^{\prime-1}(0, \alpha) \partial_{\alpha} C^\prime_{p}(0, \alpha)\\
        &+\int \frac{3d\alpha dt}{24\pi^2} \trace \left(U_p^{\prime}(0, 0, -\alpha) U_p^{\prime-1}(0, 0, t-\alpha)\right) \partial_{t}\left(U_p^{\prime}(0, 0, t-\alpha) U_p^{\prime-1}(0, 0, -\alpha)\right) C_{p}^{\prime-1}(0, \alpha) \partial_{\alpha} C^\prime_{p}(0, \alpha)\\
        &+\int \frac{3d\alpha dt}{24\pi^2} \trace \left(U_p^{\prime-1}(0, 0, t-\alpha) \partial_{t} U_p^{\prime}(0, 0, t-\alpha) U_p^{\prime-1}(0, 0, -\alpha) \partial_{\alpha} U_p^{\prime}(0, 0, -\alpha)\right)
    \end{split}
\end{equation}
The calculation yields a total of $9$ lines, which we categorize into three sets: \\
1. terms on the radius (lines $1-3$): These terms are expected to cancel each other out. This cancellation is crucial as the final outcome must remain independent of the precise definition of $U^\prime$ within the interval $0< r<1$. Provided that the conditions specified earlier are met, differing definitions of $U^\prime$ should yield identical final results.\\
2. terms on the perimeter (lines $4-6$): These terms should offset the $A$ terms in $B_{fans}$.\\
3. terms on the center (lines $7-9$): These three lines represent the final result of the calculation. Then we go through all the technical details.

First two lines vanish because $C^\prime_p(r,\alpha_p)$ is independent of $r$. We then proceed to show that the third line also evaluates to be zero under the assumption $\alpha_p=\frac{q_1}{q_2},\quad q_1,q_2 \in \mathbb{Z}$. The derivation is as follows:
\begin{equation}
    \begin{split}
        & q_1 \int 3 \trace \partial \nu \left(U_p^{\prime-1}(r, 0, t-\alpha) \partial_{\mu} U_p^{\prime}(r, 0, t-\alpha) U_p^{\prime-1}(r , 0, -\alpha) \partial_{\rho} U_p^{\prime}(r, 0, -\alpha)\right)\\
        =& q_1 \int_0^T 3 dt dr \trace \left(U_p^{\prime-1}(r, 0, t) \partial_{t} U_p^{\prime}(r, 0, t) U_p^{\prime-1}(r , 0, -\alpha_p) \partial_{r} U_p^{\prime}(r, 0, -\alpha_p)\right)\\
        =&\sum_{q=0}^{q_2-1} \int_0^{\alpha_p T} dt dr \trace U_p^{\prime-1}(r, 0, t) \partial_{t} U_p^{\prime}(r, 0, t) U_p^{\prime-1}\left(r, 0, -q \alpha_{p} T \right) C_p^q(\alpha_p) U_p^{\prime-1}\left(r,0, -\alpha_p \right) \partial_{r} U\left(r, 0, - \alpha_{p} T \right) C_p^q(\alpha_p) U\left(r, 0, - q \alpha_{p} T \right)\\
        =&\sum_{q=0}^{q_2-1} \int_0^{\alpha_p T} dt dr \trace U_p^{\prime-1}(r, 0, t) \partial_{t} U_p^{\prime}(r, 0, t)\\
        &\quad \left(U_p^{\prime-1}\left(r, 0, -(q+1) \alpha_{p} T \right) \partial_{r} U\left(r, 0, -(q+1) \alpha_{p} T \right)-U_p^{\prime-1}\left(r, 0, -q \alpha_{p} T \right) \partial_{r} U\left(r, 0, -q \alpha_{p} T \right)\right)\\
        =&0
    \end{split}
\end{equation}
The lines $4-6$ of Eq. \ref{Eq.6} is exactly $A(p,C_p)$. The contribution from lines $7-9$ of Eq. \ref{Eq.6} is derived as follows
\begin{equation}
    \begin{split}
        &\quad \int \frac{3d\alpha dt}{24\pi^2} \trace \left(U_p^{\prime}(0, 0, t-\alpha) U_p^{\prime-1}(0, 0, -\alpha)\right) \partial_{t}\left(U_p^{\prime}(1 , 0, -\alpha) U_p^{\prime-1}(0, 0, t-\alpha)\right) C_{p}^{\prime -1}(0, \alpha) \partial_{\alpha} C^\prime_{p}(0, \alpha)\\
        &+\int \frac{3d\alpha dt}{24\pi^2} \trace \left(U_p^{\prime}(0, 0, -\alpha) U_p^{\prime-1}(0, 0, t-\alpha)\right) \partial_{t}\left(U_p^{\prime}(0, 0, t-\alpha) U_p^{\prime-1}(0, 0, -\alpha)\right) C_{p}^{\prime -1}(0, \alpha) \partial_{\alpha} C^\prime_{p}(0, \alpha)\\
        &-\int \frac{3d\alpha dt}{24\pi^2} \trace \left(U_p^{\prime-1}(0, 0, t-\alpha) \partial_{t} U_p^{\prime}(0, 0, t-\alpha) U_p^{\prime-1}(0, 0, -\alpha) \partial_{\alpha} U_p^{\prime}(0, 0, -\alpha)\right)\\
        &=\frac{3}{24\pi^2}\times (2\pi)^2 \alpha_p \times \trace({H^\prime}^2(p)-2H^\prime(p) S(p) )\\
        &=\frac{1}{2}\trace({H^\prime}^2(p)-2H^\prime(p) S(p) )\\
        &=\frac{1}{2}\trace(\widetilde{H}^2(p)-S^2(p) )
    \end{split}
\end{equation}
where $\widetilde{H}(p)$ is defined as:
\begin{equation} \label{eq.H(p)}
    \widetilde{H}(p)=H^\prime(p)-S_p
\end{equation}

We sum the contributions from all three disks and obtain
\begin{equation}
\begin{aligned}
    B_{disks} &=  A\left(\Gamma, C_\Gamma(\alpha)\right)+A\left(M, M C_M(\alpha) \right)-A\left(X, C_{X}(\alpha) \right) \\
    & + \frac{1}{2}\trace \left(\widetilde{H}^2(\Gamma)-S^2_\Gamma \right)+\frac{1}{2}\trace \left(\widetilde{H}^2(M)-S^2_M \right)-\frac{1}{2}\trace \left(\widetilde{H}^2(X)-S^2_X \right)
\end{aligned}
\end{equation}
The minus sign before $X$ is because the direction of the parameter $\alpha$ is different from points $\Gamma$ and $M$, as shown by the red arrow in Fig. \ref{fig.BZ/n}.

\subsection{Merging results from previous subsections for the final results}
Combining the results of the previous two subsections, we can easily obtain the final result as follows. In Eq. \ref{Eq.7}, the integral involving $w[U]$ can be substituted by $w[U^\prime]$, differing only by an integer, since $U$ and $U^\prime$ coincide on the boundary. This relationship is expressed explicitly by:
\bea
\int_{\mathcal{L}} \frac{dk^2 dt}{24\pi^2} w[U] = B_{fans} + B_{disks} \quad \mod 1
\eea

Substituting the terms for $B_{fans}$ and $B_{disks}$ into Eq. \ref{Eq.7}, we arrive at the following expression for the final result:
\begin{equation} \label{eq.result}
        \nu = \frac{n}{2}\trace \left(\alpha_\Gamma \widetilde{H}^2(\Gamma)+\alpha_M \widetilde{H}^2(M)-\alpha_X \widetilde{H}^2(X) \right)-\frac{n}{2}\trace \left(\alpha_\Gamma S^2_\Gamma+\alpha_M S^2_M-\alpha_X S^2_X \right) \quad mod \ n
\end{equation}
The coefficients are detailed explicitly in the main text as Eq. \ref{eq.SI_2}, \ref{eq.SI_3}, \ref{eq.SI_4}, \ref{eq.SI_6}. For cases other than the $3$-fold case, $\alpha_p=\frac{1}{n_p}$. This relationship allows us to reformulate the equation as follows:
\begin{equation}
    \begin{split}
        \nu = \frac{1}{2}\trace \left(\frac{n}{n_\Gamma} \widetilde{H}^2(\Gamma)+\frac{n}{n_M} \widetilde{H}^2(M)-\frac{n}{n_X} \widetilde{H}^2(X) \right )-\frac{1}{2}\trace \left (\frac{n}{n_\Gamma} S^2_\Gamma+\frac{n}{n_M} S^2_M-\frac{n}{n_X} S^2_X \right) \quad mod \ n
    \end{split}
\end{equation}
It is important to note that the matrix $\widetilde{H}(p)$ employed here differs slightly from $\widetilde{H}_p$ as referenced in the main text; however, both matrices share identical eigenvalues. This equivalence is further elaborated upon in Appendix \ref{subsection.4}.

\section{Basis choice} 
\label{appendix.basis}
In this appendix, we explore all issues relevant to the choice of basis of Fermion states in momentum space. To provide a comparative analysis, we introduce another basis commonly used in free Fermion models, which differs from the one utilized in the proof. Consistent with Appendix \ref{appendix.proof}, we employ $a$ and $i$ as indices for sublattice and unit cell, respectively. The basis in momentum space is defined as follows:
\begin{equation} \label{eq.basis_1}
    \left |\Psi_{\vec{k},i} \right \rangle=\sum_{i} \mathbf{e}^{i\vec{k}\cdot \vec{x}_{i}}\left |\Psi_{a,i} \right \rangle
\end{equation}
The distinction of this basis from that in Eq. \ref{eq.basis} lies in the phase factor. Here, $\vec{x}_i$ represents the lattice vector of unit cell $i$, independent of the position of sublattice orbitals $a$. The advantage of this basis is continuity. For two different momentum $\vec{k}_1$ and $\vec{k}_2$ satisfying $\vec{k}_1 = \vec{k}_2 + \vec{P}_j$, where $\vec{P}_j$ is reciprocal lattice vector, the bases are the same. Explicitly,
\bea
\left |\Psi_{\vec{k}_1,i} \right \rangle = \mathbf{e}^{i \vec{P}_j \cdot \vec{x}_i}\left |\Psi_{\vec{k}_2,i} \right \rangle = \left |\Psi_{\vec{k}_2,i} \right \rangle
\eea
Consequently, this basis is defined continuously on the torus of the Brillouin zone. 

\subsection{Winding number formula}
Since the basis employed in the proof \ref{eq.basis} is not continuous on the boundary of the Brillouin zone, it is necessary to justify that the winding number, as calculated using Eq. \ref{eq.winding_number} remains valid. This validation implies that the winding number computed with this basis is the same as the winding number computed with the basis defined in Eq. \ref{eq.basis_1}. Let us denote the evolution under the basis \ref{eq.basis} and \ref{eq.basis_1} as $U$ and $U_1$ respectively. These are related trough a transformation $G$ defined in Eq. \ref{eq.G}.
\begin{equation}
U(\vec{k},t)=G(\vec{k})U_1(\vec{k},t) G^{-1}(\vec{k})
\end{equation}
Applying Eq. \ref{eq.U_212}, the winding number can be expressed as: 
\begin{equation}
    \begin{split}
        \nu[U]=\nu[U_1]
        +\int \frac{dk^2dt}{24\pi^2} \cdot 3 \trace \partial_\nu \left((U_1\partial_\mu U_1^{-1} G^{-1}\partial_\rho G)
        -(U_1^{-1}\partial_\mu U_1 G^{-1}\partial_\rho G)
        -(U_1^{-1}G^{-1}\partial_\mu G^{-1} U_1 G^{-1}\partial_\rho G)\right)
    \end{split}
\end{equation}
In this formulation, arguments of matrices are omitted for clarity. Note that even though $M(\vec{k})$ is not periodic on the torus of the Brillouin zone, $M^{-1}\partial M$ is periodic (and also momentum independent). Consequently, the boundary term vanishes because there is no boundary on the torus.

This illustration justifies the calculation of the winding number using Eq. \ref{eq.winding_number} under the basis defined in Eq. \ref{eq.basis}.

\subsection{Final result}
The final result depends on the evolution of several discrete momentum points, and the basis transformation is independent of time. The eigenvalues of matrices $S_p$ and $\widetilde{H}_p$ are invariant if they are conjugated by a unitary matrix. Therefore, the final symmetry indicator equations are universally applicable across any basis.

\section{Existence of the definition $U^\prime _p(r, \alpha=0, t)$ and $C_X^\prime(r, \alpha)$ as continuous functions of $r$}\label{appendix.tuning}
In Appendix \ref{appendix.U_prime}, we define two matrices, $U^\prime _p(r, \alpha=0, t)$ and $C_X^\prime(r, \alpha)$, and assert their existence. We further claim that the evolution matrix $U^\prime _p(r, \alpha=0, t)$ continuously depends on $r$ and satisfy boundary conditions specified in  Eq. \ref{eq.tuning} and \ref{eq.constraint}. Similarly, we assert that the matrix $C_X^\prime(r, \alpha)$ continuously depends on $r$ and satisfy the boundary conditions specified in Eq. \ref{Eq.C_X_prime} and \ref{eq.C_X_prime_condition}. In this appendix, we provide detailed proof of these claims.

\subsection{Existence of $U^\prime _p(r, \alpha=0, t)$ without the requirement that $H^\prime(p)$ commutes with $C_p$.} \label{subsection.1}
We separate the proof into two parts. The first part proofs the existence of $U^\prime _p(r, \alpha=0, t)$ without requiring that $H^\prime(p)$ commutes with $C_p$, as illustrated this in this subsection.

We first determine $U^\prime _p(r, \alpha=0, t)$ for $r \in [0, T/n_p]$ subject to the following constraint:
\bea\label{eq.U_p_prime_constraint}
\notag
&U^\prime_p(r=1,0, t)=U(\vec{k}(p),t)\\
\notag
&U^\prime_p(r=0,0, t)=\mathbf{exp}(-i 2\pi H^\prime (p) t/T)
\eea
Additionally, we require that they coincide at $t = T/n_p$:
\bea\label{eq.H_prime_U}
U(\vec{k}(p), T/n_p) = \mathbf{exp}(-i 2\pi H^\prime (p) / n_p)
\eea
This is exactly the definition that the evolution of $U(\vec{k}(p),t)$ can be continuously deformed into the evolution of $\mathbf{exp}(-i 2\pi H^\prime (p) t/T)$. This is equivalent to requiring these two evolution matrices (both as a function of $t$) to be topologically equivalent to each other. Since there exists only one topological invariant in the evolution of a $1D$ loop, namely the $1D$ winding number, the condition for equivalence is that the winding number of the following loop evolution must be zero.
\begin{equation}
U_{loop} = \left \{
\begin{array}{ll}
     U(p, t) \quad & 0 < t <  T/ n_p\\
     \exp{(i 2\pi H^\prime (p) (t/T - 1/n_p))}U(p, T /n_p)  \quad &  T /n_p < t < 2 T/n_p
\end{array} \right.
\end{equation}
This represents a loop evolution because $U_{loop}(t = 2T / n_p) = \mathbb{I}$. Writing the $1D$ winding number explicitly, we obtain the condition as follows:
\begin{equation}\label{eq.winding_1}
    \frac{i}{2\pi} \int_0^{T/n_p} dt \trace U^{-1}(t)\partial_t U(t)
    =\frac{ T}{2\pi n_p} \trace H^\prime
\end{equation}
Since the eigenvalue of $H^\prime$ are determined only up to $n_p$ by Eq. \ref{eq.H_prime_U}, we can adjust the winding number by changing eigenvalue $\lambda$ to $\lambda + n_p$. Therefore, it is always possible to find an appropriate $H^\prime(p)$ that satisfies Eq. \ref{eq.winding_1}. Thus, the existence and continuity of the evolution matrix $U^\prime _p(r, \alpha=0, t)$ are guaranteed. Specifically, $U^\prime _p(r, \alpha=0, t)$ satisfies the constraints in Eq. \ref{eq.U_p_prime_constraint} at $r = 0$ and $r=1$, while maintaining the condition $U^\prime _p(r, \alpha=0, t = T/n_p) = U(\vec{k}(p), T/n_p)$ for $r \in (0, 1)$.

The evolution matrix $U^\prime _p(r, \alpha=0, t)$ for $t> T/n_p$ can be derived using symmetry. In analogy with Eq. \ref{eq.U_Cn_matrix}, we define the matrix as follows:
\bea \label{eq.U_p_prime_symmetry}
    U_p^\prime(r, \alpha=0,t) \coloneqq C_p  U_p^\prime(r, \alpha=0,t-\frac{T}{n_p}) C_p^{-1} U_p^\prime(r, \alpha=0,\frac{T}{n_p})
\eea

This allows us to obtain the $U^\prime$ for $t\in [T/n_p, 2T/ n_p]$, and higher time intervals by induction. And the constraint $U^\prime _p(r, \alpha=0, t = T) = \mathbb{I}$ is naturally preserved under this construction.

\subsection{Existence of $U_p^\prime(r, \alpha=0,t)$ under the requirement that $H^\prime(p)$ commutes with $C_p$.}\label{subsection.2}
The second step in our proof involves establishing the existence of $U_p^\prime(r, \alpha=0,t)$ while requiring $H^\prime(p)$ to commute with the symmetry $C_p$. To achieve this, we divide the definition of $U_p^\prime(r, \alpha=0,t)$ into two parts. In the first half, i.e. $r \in [1, 1/2]$, We ensure that $U^\prime_p(r=1/2, \alpha = 0, T/n_p)$ commutes with $C_p$. And in the second part, i.e. $r \in [0, 1/2]$, we define the evolution such that $U_p^\prime(r, \alpha=0,t=0)$ is $\exp{(-i 2\pi H^\prime (p) t/T)}$.

We prebiously stated that the matrix $U_p^\prime(T/n_p, r, 0)$ is subjected to the constraint given in Eq.\ref{eq.constraint}. Substituting Eq. \ref{eq.U_p_prime_symmetry}, we obtain the following constraint:
\bea
\left ( C_p^{-1} U_p^\prime(r, \alpha =0, T/n_p) \right) ^ {n_p} = \mathbb{I}
\eea
This implies that the eigenvalues of $ C_p^{-1} U_p^\prime(T/n_p, r, 0)$ must be $\exp{(i2\pi m / n_p)}$, where $m$ is an integer. Since the manifold of the Lie group $U(n)$ is connected, we can always find a continuous path that rotates all eigenstates to eigenstates of $C_p$. As a result, $ C_p^{-1} U_p^\prime(T/n_p, r= 1/2, 0)$, and therefore $U_p^\prime(T/n_p, r= 1/2, 0)$, commutes with $C_p$. We define the path as $U_b(\beta), \beta \in [0, 1]$. The path satisfies the following conditions:
\bea
\notag
&U_b(\beta = 0) = \mathbb{I} \\
&\left [ U_b(\beta = 1) C_p^{-1} U_p^\prime(T/n_p, r= 1/2, 0) , C_p^{-1} \right ] = 0
\eea
With this construction, we can define $U^\prime_p(t, r, 0)$ for $r \in [1/2 , 1]$ as follows:
\begin{equation}
U_p^\prime(t, r, 0) = \left \{ \begin{array}{ll}
U_p^\prime(t / r , 1, 0)  & 0 < t < r T / n_p \\
C_p U_b(\beta = 2t n_p / T - 2r) C_p^{-1} U_p^\prime(T/ n_p, 1, 0)    \quad & rT / n_p < t < T/ n_p
\end{array}   \right.
\end{equation}
Note that $U^\prime_p$ at $r = 1$ is fixed by the boundary condition in Eq. \ref{eq.boundary_condition}. This definition satisfies the requirement that $U^\prime_p(r=1/2, \alpha = 0,T/n_p)$ commutes with $C_p$. 

The second part, for $r \in [0, 1/2]$, is to define the evolution such that when $r = 0$, 
\bea
&U_p^\prime(r = 0, \alpha = 0, t) = \exp{(-i 2\pi H^\prime (p) t)} \\
&\exp{(-i 2\pi H^\prime (p) /n_p)} = U_p^\prime(r = 1/2, \alpha = 0, t = T/n_p) \label{eq.bc_1}
\eea
Following the last subsection \ref{subsection.1}, the Eq. \ref{eq.bc_1} ensures that we can find an $H^\prime(p)$ that commutes with $C_p$ while also satisfying the condition given in Eq. \ref{eq.1d_winding}. This concludes the proof of the existence. 

\subsection{Subtlety about point $X$}\label{subsection.3}
The existence of $C_X^\prime(r, \alpha)$ satisfying the conditions in Equations \ref{Eq.C_X_prime} and \ref{eq.C_X_prime_condition} has been proved in Subsection \ref{subsection.1} for all cases except the $3$-fold case. The parameter $t$ can be changed into $\alpha$ and proof will apply. However, there are specific subtleties regarding the $3$-fold case and we will address in this subsection.

For point $X$, we defined $C_X(\alpha)$ for $\alpha \in [0, \alpha_p T] = [0, 2 / 3]$, while the result in the main text is for the rotation by angle $1/3$ times $2\pi$. (For all the other cases, $\alpha_p = 1/ n_p$.) To prove the existence of $C_x^\prime(r, \alpha)$, we need to demonstrate the following:
\bea
\frac{i}{2\pi} \int_0^{2/3} d \alpha \trace C_X^{-1}(\alpha)\partial_\alpha C_X(\alpha) = \frac{i}{2\pi} \int_0^{2/3} d \alpha \trace \mathbf{exp}(i 2\pi S_X \alpha)\partial_\alpha \mathbf{exp}(-i 2\pi S_X \alpha)
\eea
Substituting the definition of $C_X(\alpha)$ from Eq. \ref{eq.C_X_alpha}, this condition becomes:
\bea
\trace S_\Gamma / 3 + \trace S_M / 3 = 2 \trace S_X / 3
\eea
Although this condition may not be satisfied by adjusting the eigenvalues of $S_X$ due to the factor of $2$, it can be fulfilled by adjusting the eigenvalues of $S_\Gamma$ or $S_M$, specifically by changing the eigenvalue by 3.  We reiterate that this adjustment of $S_p$ does not affect the final result, as $S_p$ appears in both $S_p$ and $\widetilde{H}_p$.

\subsection{Rewrite the condition with $\widetilde{H}$} \label{subsection.4}
In the final result, only $\widetilde{H}$ appears, which is why we focus solely on $\widetilde{H}$ in the main text. It is therefore essential to reformulate the condition in Eq. \ref{eq.winding_1} that the evolution $U(\vec{k}(p),t)$ can be continuously changed to the evolution $\exp{(-i 2\pi H^\prime (p) t/T)}$ in terms of $\widetilde{H}$. To do this, we apply the symmetry operation $C_p$ to the end of both the evolution matrices.  The evolution $U(\vec{k}(p),t)$ is changed to Eq. \ref{eq.U_0} in the main text, while $\exp{(-i 2\pi H^\prime (p) t/T)}$ is changed to $\exp{(-i 2\pi \widetilde{H} (p) t/T)}$, because $C_p$ commutes with $H^\prime(p)$. Thus, our condition in Eq. \ref{eq.winding_1} in terms of $\widetilde{H}_p$ is that there exists a continuous function $\widetilde{U}_p(r, t)$ with $r\in [0, 1]$, such that:
\bea
&\widetilde{U}_p(r=1, t) = \widetilde{U}_0(p, t) \\
&\widetilde{U}_p(r=0, t) = \exp{(-i 2\pi \widetilde{H}_p t /T)}
\eea
where $\widetilde{U}_0(p, t)$ is defined in the main text Eq. \ref{eq.U_0}, and $\widetilde{H}_p$ is time-independent. We do not have to require $\widetilde{H}_p$ to commute with $C_p$ because the second part of $U^\prime_p$ written in subsection \ref{subsection.2} does not change the eigenvalues. So, this $\widetilde{H}_p$ has the same eigenvalue as $\widetilde{H}(p)$. And $\widetilde{H}(p)$ commutes with $S_p$ based on its definition in Eq. \ref{eq.H(p)}.

And the condition for this tuning is changed from Eq. \ref{eq.winding_1} to Eq. \ref{eq.1d_winding}.

To easily determine the eigenvalues of $\widetilde{H}_p$, we note that the formula for the $1D$ winding number depends only on the eigenvalues, as expressed by
\begin{equation}
\begin{split}
\nu[U] &= \frac{i}{2\pi}  \int  dt\trace U^{-1} \partial_t U  = \frac{i}{2\pi} \sum_j \int dt \lambda_{U,j}^{-1} \partial_t \lambda_{U,j} \\
& = \frac{1}{2\pi} \sum_j \int dt \partial_t \phi_{U, j} = \sum_j \frac{1}{2\pi} (\phi_{U, j}(t_2) - \phi_{U, j}(t_1))
\end{split}
\end{equation}
where $\lambda_{U, j} = \exp{(i\phi_{U, j})}$ is the $j$-th eigenvalue of matrix $U$, and $t_1, t_2$ are the upper and lower bounds of the integral. By plotting the phase of the eigenvalues $\phi_{U, j}(t)$ as a function of time, the winding number can be directly identified. In the main text, we use this method to determine the eigenvalues of $\widetilde{H}_p$.

It is also important to highlight that all the requirements in this section involve the evolution operator $U, \widetilde{U}_0, U_p^\prime \cdots$. In the final result, the specific definition of symmetry operator $S_p$ does not affect the outcome, as long as it satisfies Eq. \ref{eq.S_p}. Using a property of $1D$ winding number,
\bea
\int  dt\trace (U_1U_2)^{-1} \partial_t (U_1 U_2)  = \int  dt \trace U_1^{-1} \partial_t U_1 + \int  dt\trace U_2^{-1} \partial_t U_2 
\eea
Thus, if we redifine the eigenvalue of $S_p$ from $\lambda$ to $\lambda + n_p$, the eigenvalue of $\Tilde{U}_0(p, t)$ will also change by $n_p$, ensuring that the result remains unchanged, at least modulo $n$.

\section{Reduced Brillouin zone of $2$-fold case different from the other three cases}\label{appendix.C2}
\begin{figure}[h]
    \centering
    \includegraphics[width=0.3\columnwidth]{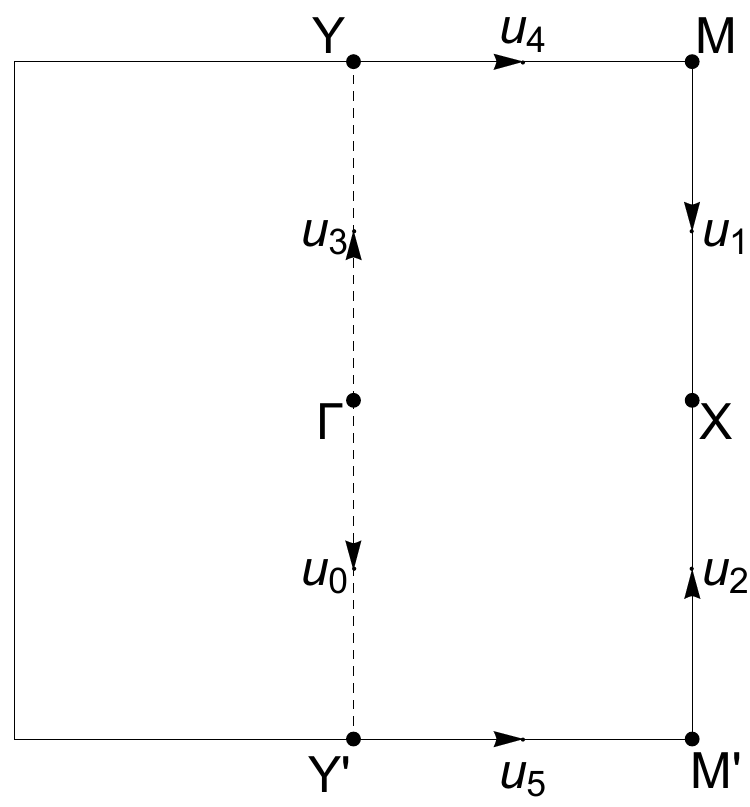}
    \caption{reduced Brillouin Zone $\mathcal{L}$ for $C_2$ time screw symmetry. Different from the $3$, $4$ and $6$ fold case, the edges of this region is divided into $6$ segments by symmetric points.}
    \label{fig.BZ2}
\end{figure}

In $2$-fold symmetric systems, The boundary of region $\mathcal{L}$ consists six line segments, as presented in Fig. \ref{fig.BZ2}, grouped into three pairs, each related by symmetry. 
\begin{equation}
    \begin{split}
        \vec{k}_3& =\hat{C}_2 \vec{k}_0\\
        \vec{k}_1& =\hat{C}_{2} \vec{k}_2+\vec{P}_x \\
        \vec{k}_4& =\vec{k}_5 + \vec{P}_y  
    \end{split}
\end{equation}

We follow our previous proof with minor changes. In Eq. \ref{Eq.8}, the terms involving $(u_0,u_3)$ and $(u_1,u_2)$ are the same as before. But besides that, additional terms involving $(u_4,u_5)$ appear as follows:
\begin{equation}
    \begin{split}
        &+\frac{3}{24\pi^2}\int_0^\pi dk\int dt\  \trace ( U^{-1}(\vec{k}_4,t)\partial _t U(\vec{k}_4,t) U^{-1}( \vec{k}_4,-\frac{T}{2})\partial_x U( \vec{k}_4,-\frac{T}{2}))\\
        &-\frac{3}{24\pi^2}\int_0^\pi dk\int dt\  \trace ( U^{-1}(\vec{k}_5,t)\partial _t U(\vec{k}_5,t) U^{-1}( \vec{k}_5,-\frac{T}{2})\partial_x U( \vec{k}_5,-\frac{T}{2}))      
    \end{split}
\end{equation}
For the same reason as in Eq. \ref{Eq.9}, these two terms are related by a gauge transformation $G_y$ and thus cancel each other. So no boundary terms involving $(u_4,u_5)$ contribute to Eq. \ref{Eq.7}. In the next step, when constructing the unitary transformation in Eq. \ref{Eq.10}, three fans are required instead of two, with the third one corresponding to $u_4,u_5$. Since there is no time translation between $u_4$ and $u_5$, we use the following equation to replace Eq. \ref{Eq.10}
\begin{equation}
    U_{s}^{\prime}(k, \alpha, t)=G_y^{-1}(\alpha) U(\vec{k}_{s}, t)  G_y(\alpha)
\end{equation}
Thus, no term involving $\partial_\alpha$ appears in Eq. \ref{Eq.11}. Up to now, we have explained why the terms relevant to boundary $u_4, u_5$ vanish. 

The remaining steps proceed as in the previous case. We now have $4$ disks of $U^\prime$ corresponding to $4$ different high-symmetry points $\Gamma, M, X, Y$. the final result is similar to Eq. \ref{eq.result}, but now incorporates these four points:
\begin{equation}
\begin{split}
        \nu &= \frac{n}{2}\trace \left(\alpha_\Gamma \widetilde{H}^2(\Gamma)+\alpha_M \widetilde{H}^2(M)-\alpha_X \widetilde{H}^2(X) - \alpha_Y \widetilde{H}^2(Y) \right)-\frac{n}{2}\trace \left(\alpha_\Gamma S^2_\Gamma+\alpha_M S^2_M-\alpha_X S^2_X - \alpha_Y S^2_Y \right) \quad mod \ n \\
        &= \frac{1}{2} \trace \left( \widetilde{H}^2(\Gamma)+\widetilde{H}^2(M)-\widetilde{H}^2(X) -  \widetilde{H}^2(Y) \right)-\frac{n}{2}\trace \left(S^2_\Gamma+ S^2_M- S^2_X - S^2_Y \right) \quad mod \ 2
\end{split}
\end{equation}
where in the second line, we substitute $n = 2$ and set all the $\alpha$ values equal to $1/2$.

\section{Symmetry indicators of $C_n$ rotational symmetry} \label{appendix.rotation}
Thus far, we have focused on time screw symmetry. The same method extends naturally to rotational symmetry, with a simpler proof. For a system exhibiting $C_n$ space rotational symmetry, the Hamiltonian satisfies the relation:
\begin{equation}\label{rotational_symmetry}
    \hat{H}(\hat{C}_n \vec{k},t)=C_n \hat{H}(\vec{k},t) C_n^{-1}
\end{equation}
We can also employ the basis with constant $C_n$ and express the total winding number as evolution in the reduced Brillouin zone $\mathcal{L}$
\begin{equation}
    \nu[U]=n \int_{\mathcal{L}} w[U]\frac{dk^2 dt}{24\pi^2}
\end{equation}
The next step involves replacing $U$ with $U^\prime$ to explicitly calculate $w[U]$. We utilize 
\begin{equation}
U_{s}^{\prime}(k, \alpha, t)=C(\alpha) U(\vec{k}_{s}, t) C^{-1}(\alpha)
\end{equation}
Thus, the winding number density of this evolution is 
\begin{equation}
\begin{split}
&\int \frac{dtdkd\alpha}{24\pi^2} w\left[U_{s}^{\prime}(k, t, \alpha)\right]\\
&=3 \int \frac{dtdkd\alpha}{24\pi^2} \partial_{k}(U(\vec{k}_s, t) \partial_{t}U^{-1}(\vec{k}_{s},t) {C}^{-1}(\alpha) \partial_{\alpha} C(\alpha))\\
&-3 \int \frac{dtdkd\alpha}{24\pi^2} \partial_{k}(U^{-1}(\vec{k}_{s}, t) \partial_{t}U(\vec{k}_{s}, t) C^{-1}(\alpha) \partial_{\alpha} C(\alpha)) 
\end{split}
\end{equation}
For simplicity, we define the following function:
\begin{equation}
\begin{split}
A(p, C) &\equiv \frac{3}{24\pi^2} \int_0^T d \alpha d t  \trace\\
&\quad U(\vec{k}_s, t) \partial_{t}U^{-1}(\vec{k}_{s},t) {C}^{-1}(\alpha) \partial_{\alpha} C(\alpha))\\
&-U^{-1}(\vec{k}_{s}, t) \partial_{t}U(\vec{k}_{s}, t) C^{-1}(\alpha) \partial_{\alpha} C(\alpha) \\
\end{split}
\end{equation}
to rewrite the winding number density as
\begin{equation} \label{Eq.13}
    \begin{split}
        \sum_{s=0,2}\int \frac{dtdkd\alpha}{24\pi^2} w\left[U_{s}^{\prime}(\vec{k}_{s}, \alpha, t)\right]
         =& -A\left(\Gamma, C_\gamma(\alpha) \right)-A\left(M, C_M (\alpha) \right)+A\left(X, C_\Gamma(\alpha) \right)+A\left(Y , C_M(\alpha)\right)\\
         =& -A\left(\Gamma, C_\Gamma(\alpha)\right)-A\left(M, C_M(\alpha)\right)+A\left(X, C_X(\alpha)\right)
    \end{split}
\end{equation}
where the second line holds for the same reasons as previously discussed, as definition of $C_p$ remains unchanged. The next step is to define another $U^\prime_p(r,\alpha, t)$ on the disks to fill in the holes. However this time, the holes at the points $\Gamma$ and $M$ are already filled because 
\begin{equation}
     U_s^\prime (k=0,\alpha,t)=C ( \alpha ) U ( \vec{k} (\Gamma) ,t ) C^{-1} ( \alpha ) =U ( \vec{k} (\Gamma) ,t )
\end{equation}
 is independent of $\alpha$ due to symmetry. However, for the hole in the middle, we still need to define
 \begin{equation}
     U^\prime_X(r,\alpha, t)= C^\prime_p(r,\alpha) U^\prime_X(r,0, t) C_p^{\prime-1}(r,\alpha)
 \end{equation}
This time, there is no need to adjust $U^\prime_X(r, 0, t)$, meaning $U^\prime_X(r,0, t) \coloneqq U^\prime_X(r=1,0, t)$. Only $C_p(r,\alpha)$ is dependent on $r$ according to Eq. \ref{Eq.C_X_prime}
\begin{equation}\label{Eq.14}
    \begin{split}
        C^\prime_X(r=1,\alpha) &= C_X(\alpha)\\
        C^\prime_X(r=0,\alpha) &= \mathbf{exp}(-i 2\pi S_X \alpha)
    \end{split}
\end{equation}
Performing a similar calculation to Eq. \ref{Eq.6}, we obtain:
\begin{equation}
    \begin{split}
        w[U^\prime_X(r,\alpha, t)]&= \int \frac{3d\alpha dt}{24\pi^2} \trace U^\prime_X(t,r=1,0)\partial_t U_X^{\prime-1}(r=1,0, t) C_X^{\prime-1}(r=1,\alpha) \partial_\alpha C^\prime_X(r=1,\alpha)\\
        &-\int \frac{3d\alpha dt}{24\pi^2} \trace  U_X^{\prime-1}(r=1,0, t)\partial_t U^\prime_X(r=1,0, t) C_X^{\prime-1}(r=1,\alpha) \partial_\alpha C_X^\prime(r=1,\alpha)\\
        &- \int \frac{3d\alpha dt}{24\pi^2} \trace U^\prime_X(r=0,0, t)\partial_t U_X^{\prime-1}(r=0,0, t) C_X^{\prime-1}(r=0,\alpha) \partial_\alpha C_X^\prime(r=0,\alpha)\\
        &+\int \frac{3d\alpha dt}{24\pi^2} \trace U_X^{\prime-1}(r=0,0, t)\partial_t U^\prime_X(r=0,0, t) C_X^{\prime-1}(r=0,\alpha) \partial_\alpha C^\prime_X(r=0,\alpha)\\
        &= A(X,C_X(\alpha))-A(X,C^\prime_X(r=0,\alpha))
    \end{split}
\end{equation}
This implies that we can also replace the evolution $C_X(\alpha)$ by a constant evolution of $\exp{(-i2\pi S_x \alpha)}$ in Eq. \ref{Eq.13}. Each term in Eq. \ref{Eq.13} can be calculated as:
\begin{equation}
    \begin{split}
        A(p,C_p)=&\frac{-3i\times 2\pi \alpha_p}{24 \pi^2}\int_0^T dt \trace (U(\vec{k}_s,t)\partial_t U^{-1}(\vec{k}_s,t) S(p)+U(\vec{k}_s,t)\partial_t U^{-1}(\vec{k}_s,t) S(p))\\
        =&\frac{-i\alpha_p}{2\pi}\int_0^T dt \trace (U^{-1}(\vec{k}_s,t)\partial_t U(\vec{k}_s,t) S(p))
    \end{split}
\end{equation}
where we used the fact that $U$ commutes with $C_p$, and thus with $S_p$. Consider the winding number in  one (time) dimensional
\begin{equation}
    \nu[U(t)]=\frac{i}{2\pi}\int dt \trace (U^{-1}\partial_t U)
\end{equation}
In this case, the winding number corresponds to different bands, each multiplied by the associated  symmetry quantum number. We denote this as:
\begin{equation}
    \nu_{n_p} (p)=\frac{i}{2\pi}\int dt \trace (U^{-1}\partial_t U S_p)
\end{equation}
where $C_p$ is the symmetry for point $p$, and $S_p$ is logarithm of $C_p$, defined by:
\begin{equation}
    \mathbf{e}^{-i 2 \pi S_p /n_p} = C_{p}
\end{equation}
Thus, the final result is: 
\begin{equation}
\nu=n \alpha_\Gamma \nu_{n_\Gamma}(\Gamma)+n \alpha_M \nu_{n_M}(M)-n \alpha_X \nu_{n_X}(X) \quad mod \ n
\end{equation}
For every cases ($C_2,C_3,C_4,C_6$), we can write it explicitly
\begin{equation}
    \begin{split}
        \nu=& \nu_2(\Gamma)+\nu_2(M)-\nu_2(X)-\nu_2(Y) \quad mod \ 2\\
        \nu=& \nu_3(\Gamma)+\nu_3(M)-2\nu_3(X) \quad mod \ 3\\
        \nu=& \nu_4(\Gamma)+\nu_4(M)-2\nu_2(X) \quad mod \ 4\\
        \nu=& \nu_6(\Gamma)+2\nu_3(M)-3\nu_2(X) \quad mod \ 6
    \end{split}
\end{equation}
By using the property $\nu_n \in \mathbb{N}$, we can replace all the negative signs with positive ones. The resulting expressions are
\begin{equation}
    \begin{split}
        \nu=& \nu_2(\Gamma)+\nu_2(M)+\nu_2(X)+\nu_2(Y) \quad mod \ 2\\
        \nu=& \nu_3(\Gamma)+\nu_3(M)+\nu_3(X) \quad mod \ 3\\
        \nu=& \nu_4(\Gamma)+\nu_4(M)+2\nu_2(X) \quad mod \ 4\\
        \nu=& \nu_6(\Gamma)+2\nu_3(M)+3\nu_2(X) \quad mod \ 6
    \end{split}
\end{equation}
This can be summarized as 
\bea
\nu = \sum_p{\frac{n}{n_p} \nu_{n_p}(p)} \quad \mod n
\eea
where $p$ is summed over all the independent high-symmetry points in the Brillouin zone.

\section{General gapped Floquet systems} \label{appendix.gap}
\begin{figure}[h]
    \centering
    \includegraphics[width=0.4 \columnwidth]{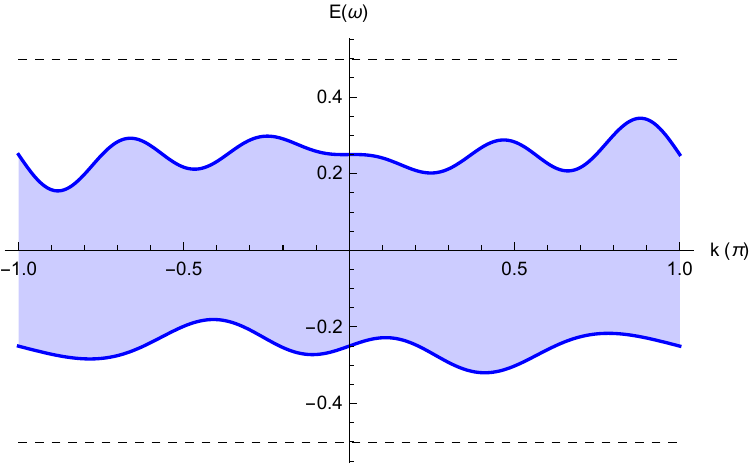}
    \includegraphics[width=0.4 \columnwidth]{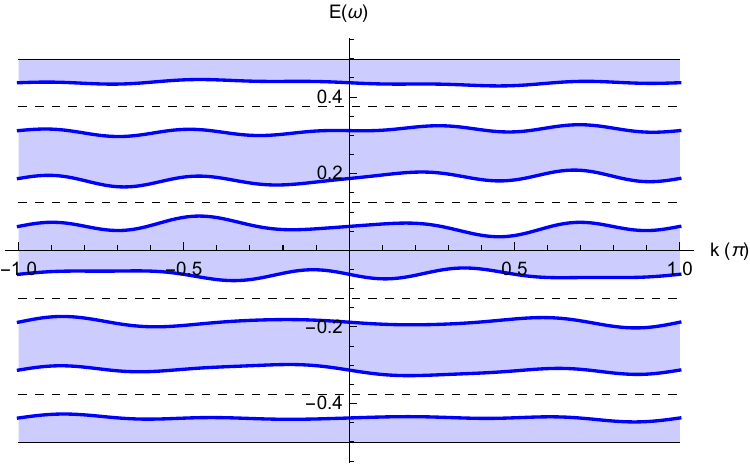}
    \caption{Schematic diagram of the spectrum for general gapped bands of Floquet evolution $U(T)$ and $C_4^{-1}U(T/4)$. If $U(T)$ shows gap at $\frac{\omega}{2}$, $C_4^{-1}U(T/4)$ will exhibit a gap at $\omega \left(\frac{1}{8} + \frac{m}{4} \right)$. The blue continuum represent the bands and grey dashed lines indicate the gaps.}
    \label{fig.band}
\end{figure}

Thus far, we have only considered the case of $U(\vec{k}, 2\pi)=\mathbb{I}$, which is not general enough for practical purposes. It is well-known that topological phases can be defined for more general evolution as long as they remain gapped, and the same applies to symmetry indicators. The quasi-energy spectrum of a unitary operator is defined as $E = \frac{-i}{T} \ln{\lambda}$, where $\lambda$ is an eigenvalue of the unitary evolution matrix. In Fig.\ref{fig.band}, we show a general band structure with gap at $E=\omega/2$, indicated by dashed lines. Due to time screw symmetry, we have the following relation:
\bea
U(\vec{k}(p), 2\pi) = \left ( C_p^{-1} U(\vec{k}(p), 2\pi/n_p) \right )^{n_p} 
\eea
Thus, the operator $\widetilde{U}_p$ must be gapped at $E=\omega \left (\frac{1}{2n_p}+\frac{m}{n_p} \right), m\in \mathbb{Z}$.
If we continuously tune the spectrum of $U(T)$  to 0, the spectrum of $U_0$ will converge to flat bands at $\frac{m\omega}{n_p}$ for all the states within the range $\omega(\frac{m}{n_p} \pm \frac{1}{2n_p})$. Since only the eigenvalues are relevant for the calculation of symmetry indicators, we can still define and compute the eigenvalues of  $\widetilde{H}_p$. The eigenvalues are not necessarily integers for a general $U(\vec{k}, t)$, but we can simply round to the nearest integer. For example, we select $n$ if eigenvalue is within the interval $[n-0.5, n+0.5]$, ensuring that our symmetry indicator formula remains valid.

\end{document}